\documentclass{article}
\begin{document}

\title{`Majorana mass' fermions as untrue 
Majorana particles, rather endowed with 
pseudoscalar-type charges than genuinely neutral} 

\author{G. Ziino\\Dipartimento di Fisica e Chimica,\\
Universit\`a di Palermo, via Archirafi 36, I - 90123 Palermo (Italy); 
\\ e-mail address: giorgio.ziino@unipa.it ; fax:  +39-091-6234232}

\date{}

\maketitle

\begin{abstract}
The idea of a `Majorana mass' to make a chiral neutrino 
really neutral is here reconsidered. It is pointed out 
that such an approach, unlike Majorana's (non-chiral) 
old one, does not strictly lead, in general, to a true 
self-conjugate particle. This can be seen on directly 
using the basic definition (or fundamental representation) 
of charge conjugation $C$ in Quantum Field Theory, as an 
operation just acting on annihilation and creation operators 
and just expressing particle--antiparticle interchange. 
It is found, indeed, that the `active' and `sterile' 
whole fields which can be obtained from mixing the chiral 
components of two mutually charge-conjugate Dirac fields 
are themselves `charge conjugate' to each other (rather 
than individually self-conjugate). These fields, 
taken as mass eigenfields (as in the `Majorana mass' 
case), are shown to describe particles carrying 
pseudoscalar-type charges and being neutral relative 
to scalar-type charges only. For them, `$CP$ symmetry' 
would be nothing but pure mirror symmetry, and $C$ 
violation (already implied in their respective `active' 
and `sterile' behaviors) should then involve time-reversal
violation as well. The new (no longer strictly chargeless) 
`Majorana mass' neutrino model still proves, however, 
neither to affect the usual expectation for a neutrinoless 
double $\beta$-decay, nor to prevent `active' and `sterile' 
neutrino varieties from generally taking different 
mass values. One has, on the other hand, that any fermion 
being just a genuine (i.e. really self-conjugate) 
Majorana particle cannot truly exist in two distinct 
$-$ `active' and `sterile' $-$ versions, and it can further 
bear only a unified mass kind which may at once be said 
to be either a `Majorana-like' or a `Dirac-like' mass kind. 
\end{abstract}


\section{Introduction}
The `nature' of the neutrino mass (if any) 
has been for years one of the most intriguing 
puzzles of elementary-particle physics. 
The point still at issue is whether real neutrinos 
may even be {\it self-conjugate} (or self 
charge-conjugate) like Majorana particles 
$[{\ref{Majorana1937}}]$, and further endowed 
with so-called `Majorana masses' $[{\ref{Jehle1949},
\ref{Serpe1949}}]$, or there actually exist mere 
`Dirac mass' neutrinos looking like standard 
fermions ({\it different} from their own 
antiparticles). In this regard, the most general 
neutrino model available for each lepton family 
is believed to include an overall Lagrangian 
mass term with both `Dirac' and `Majorana' 
contributions, and with the latter contribution 
being the sum of two distinct mass terms for 
wholly `active' and `sterile' neutrino types 
$[{\ref{Esposito1998}}]$. Such a model, if 
suitably conceived with a `Majorana' sector 
having just one nonzero term, relevant to 
a super-heavy `sterile' neutrino type, 
can in particular be seen to account $-$ 
via the well-known See--Saw Mechanism  
$[{\ref{Gell-Mann1979}}]$ $-$ for the 
very small size to which the actual 
neutrino masses seem to be confined. 
Renewed interest in Majorana's conjecture 
has been recently aroused also by 
the experimental discovery of Majorana 
bound states (or quasiparticles) in 
superconductors $[{\ref{Mourik2012}}-{\ref{Esposito2013}}]$.

	This paper deals with some subtle, 
and not yet thoroughly investigated, basic 
theoretical aspects concerning the idea 
itself of a `Majorana' Lagrangian mass term 
as a way to get either an `active' or a 
`sterile' up-to-date version of the 
original (manifestly self-conjugate) 
Majorana neutrino field. It should, first 
of all, be reminded that the `Majorana mass' 
construct cannot be really traced back to 
Majorana himself, nor can it be said essential 
for a self-conjugate fermion. Despite this, 
that a `Majorana mass' fermion should just be 
a Majorana particle $-$ i.e. a really 
neutral fermion $-$ is normally regarded as 
quite an obvious conclusion, which automatically 
follows from an extended use of the well-known 
formula $-$ Eq.~(\ref{2.9bis}) $-$ defining 
the `charge conjugate' of a standard 
Dirac field. According to common views, 
the full legitimacy of such a procedure 
is in particular believed to be unquestionable. 
Doing like that, however, one is {\it not} 
directly applying charge conjugation (or 
particle--antiparticle conjugation) as it 
is {\it primarily} defined within Quantum 
Field Theory (QFT): namely, an operation, 
$C$, truly acting on annihilation and 
creation operators and merely consisting 
in turning them into their own `charge conjugates' 
(with no changes in either four-momenta or 
helicities) $[{\ref{Merzbacher1970}}]$. 
As already pointed out by Dvornikov 
in his canonical quantization of a massive 
Weyl field $[{\ref{Dvornikov2012}}]$, 
this should not be taken as a negligible detail. 
One may guess it even better on going over 
to the zero-mass case. It is sufficient, 
for example, to take account of the 
straightforward hints here given in Sec.~2, 
on how to interpret the two couples of 
fermionic and antifermionic Weyl solutions 
in order to make sure that $C$ may really have
{\it no} effects on helicities. These hints 
show that an approach like the usual one 
in defining the `charge conjugate' of a 
Weyl field does not seem at all to lead 
to the appropriate choice. They also 
suggest the need for a more general 
check on the real consistency of 
a procedure that does nothing but 
{\it borrow} the standard definition 
of a `charge conjugate' Dirac field. 
The simplest way to do so is just to make 
{\it direct} use of the above-mentioned 
fundamental representation of $C$. 
Following this way, a basic {\it new} outcome 
is here obtained which overturns the current 
reading. It is found, indeed, that an 
{\it active} `Majorana mass' fermion field 
and its {\it sterile} counterpart prove 
rather to be {\it mutually charge-conjugate} 
than individually self-conjugate, and so, 
at most, they may give rise to {\it one 
and the same} (really neutral) Majorana 
field if they are further imposed to coincide.
 
	The formal aspects relevant to the whole 
question (including an explicit representation 
of the effective `new' action of $C$ on single 
chiral fields) are widely discussed in Sec.~3. 
In the subsequent section, moreover, it is shown 
that, regardless of whether the `Majorana mass'
or `Dirac mass' case is considered, there 
generally exists a fully {\it symmetrical} 
link connecting both an active spin-$\frac{1}{2}$ 
field and its {\it charge conjugate} sterile 
counterpart with the corresponding pair of 
{\it charge conjugate} Dirac fields whence 
they have been constructed. This link is given 
by a unitary transformation being as well 
the {\it inverse} of itself, and it suitably 
allows an {\it extended} ($8\times8$) matrix 
representation for $C$. In the light of the 
new formalism, on the other hand, the conclusion 
may also be drawn that a {\it true} (really 
self-conjugate) Majorana field can no longer 
turn out to be of two {\it different} $-$ 
`active' and `sterile' $-$ types, and furthermore 
(in strict accordance with its self-conjugate 
nature) it can be assigned only a {\it unified} 
mass kind which may at once be viewed as 
either a `Majorana-like' or a `Dirac-like' 
mass kind.

	In Secs.~5, 6 and 7, a full insight 
is gained into the general variety of `charges' 
that should now characterize a genuine 
`Majorana mass' fermion and tell it from 
a genuine `Dirac mass' fermion. The former 
particle, unlike the latter, should actually 
be endowed with {\it pseudoscalar-type} 
(or {\it axial-type}) charges and be `neutral' 
as regards {\it scalar-type} charges only 
[{\ref{Dvoeglazov2012}}]. The `neutrality'
of it, in other words, is now to be meant
no longer under $C$, but rather under a
{\it more restrictive} `charge conjugation' 
operation which leaves pseudoscalar-type 
charges unvaried and merely corresponds to 
a `scalar-charge conjugation' operation. 
One such charged spin-$\frac{1}{2}$ 
particle would {\it in turn} amount to 
a `fermion' or an `antifermion' depending 
on {\it either} chirality involved. Thus, 
for instance, an active `Majorana mass' 
neutrino is to be now referred to as a `lepton' 
(having {\it positive} lepton number) or 
an `antilepton' (having {\it negative} 
lepton number) according to whether being 
a {\it left-handed} or a {\it right-handed} 
particle, whereas the exact converse (with 
`lepton' and `antilepton' {\it interchanged}) 
should hold for the `charge conjugate' 
sterile counterpart of it. In close 
connection with this, active and sterile 
`Majorana mass' neutrinos may now be 
regarded as truly {\it obeying} ordinary 
mirror symmetry as just the analogue of 
`$CP$ symmetry' for Dirac neutrinos. 
A manifest (maximum) $C$ violation is 
to be instead recognized in their 
(maximally) asymmetrical dynamical 
behaviors, and this should actually imply, 
in the light of the $CPT$ theorem, a 
(maximum) {\it time reversal} violation 
as well (just counterbalancing the 
`recovered' $P$ symmetry). The new 
reading can also be seen, in particular, 
neither to influence the usual expectation 
for a neutrinoless double $\beta$-decay, 
nor to rule out the possibility $-$ still 
compatible with $CPT$ symmetry $-$ of  
{\it different} mass values for the two 
(active and sterile) `Majorana mass' 
neutrino versions. 

	In Sec.~8, finally, it is pointed out 
that a pair of charge conjugate spin-$\frac{1}{2}$ 
fields with identical masses, whether it 
may be a `Dirac mass' or a `Majorana mass' 
field pair, can always be expressed as a linear 
combination of a couple of {\it true} Majorana 
fields with opposite $CP$ intrinsic parities 
(and identical masses).

\section{A `Majorana mass' neutrino as not exactly 
a genuine (really neutral) Majorana particle}
	According to Majorana's early approach 
$[{\ref{Majorana1937}}]$, a {\it self-conjugate} 
neutrino is a really neutral spin-$\frac{1}{2}$ 
particle which may be formally assigned, say, 
a Dirac field solution of the special type
\begin{equation}
\psi_{\rm M}(x) = \frac{1}{\sqrt{2}}\left[\psi(x) + \psi^c(x)\right]
\label{2.1}
\end{equation}

\noindent
($x \equiv x^\mu; \mu=0,1,2,3$), where $\psi^c(x)$,
defined as
\begin{equation}
\psi^c(x) \equiv C \psi(x) C^{-1}
= U_C\psi^{\dagger {\rm T}}(x),
\label{2.9bis}
\end{equation}

\noindent
is the charge conjugate of a standard Dirac 
field solution $\psi(x)$, such that $\psi(x)\not=\psi^c(x)$. 
Here $U_C$ denotes the usual charge-conjugation (or $C$) 
matrix, and $\psi^{\dagger {\rm T}}$ is the transpose 
of the adjoint solution $\psi^\dagger$. The 
field given by Eq.~(\ref{2.1}) has a {\it manifest} 
self-conjugate form: 
\begin{equation}
\psi_{\rm M}^c(x) = \frac{1}{\sqrt{2}}\left[\psi^c(x) + \psi(x)\right]
=  \psi_{\rm M}(x),
\label{2.1bis}
\end{equation}

\noindent
and it can thus be automatically expanded in terms
of {\it net} annihilation and creation operators 
coinciding with their own charge conjugates.
This field, in other words, is self-conjugate  
{\it by definition}; and the associated fermion, 
usually known as a {\it Majorana particle}, 
is such that it cannot possibly be distinguished 
from its antiparticle. If we in particular split 
$\psi_{\rm M}$ into a {\it left-handed} 
chiral component, $\frac{1}{2}(1 - \gamma^5)
\psi_{\rm M}$, plus a {\it right-handed} one, 
$\frac{1}{2}(1 + \gamma^5)\psi_{\rm M}$, where 
$\gamma^5$ ($\equiv i\gamma^0\gamma^1\gamma^2\gamma^3$) 
is just denoting the chirality matrix, we then have 
that the former (latter) component taken alone 
would indifferently be able to describe a left-handed 
{\it neutrino} (right-handed {\it antineutrino}) 
as well as a left-handed {\it antineutrino} 
(right-handed {\it neutrino}):
\begin{equation}
\frac{1}{2}(1 \mp \gamma^5)\psi_{\rm M} =
\frac{1}{2}(1 \mp \gamma^5)\psi_{\rm M}^c.
\label{2.2bis}
\end{equation}
 
\noindent
In this regard, it is worth pointing out that 
$\psi_{\rm M}$, also expressible {\it in the form}
\begin{equation}
\psi_{\rm M}=\frac{1}{\sqrt{2}}\left[\frac{1}{2}(1 - \gamma^5)\psi
\!+\! \frac{1}{2}(1 + \gamma^5)\psi^c\right] 
+ \frac{1}{\sqrt{2}}\left[\frac{1}{2}(1 + \gamma^5)\psi
\!+\! \frac{1}{2}(1 - \gamma^5)\psi^c\right],
\label{2.1ter}
\end{equation}

\noindent 
can by no means be assigned any special sort 
of `handedness' marking it as {\it either} 
an `active' {\it or} a `sterile' field. 
As shown by (\ref{2.1ter}), one strictly 
has that `active' and `sterile' contributions 
are always {\it equally present} in $\psi_{\rm M}$\,! 
This, indeed, implies that a neutrino just described 
by a field like $\psi_{\rm M}$ would not be compatible
with the Standard Model (SM) $[{\ref{Glashow1961}
-\ref{Salam1968}}]$, as it could give only {\it half} 
of the required contribution to the square modulus 
of the matrix element. 

	The equation obeyed by $\psi_{\rm M}$ 
(still the Dirac one) may be derived, as usual, 
from a free spin-$\frac{1}{2}$ quantum field 
Lagrangian with a mass term proportional to
\begin{equation}
\bar {\psi}_{\rm M}\psi_{\rm M},
\label{2.1quater}
\end{equation}

\noindent
where $\bar {\psi}_{\rm M} = \psi_{\rm M}^\dagger\gamma^0$.
This term looks just like the one for a standard Dirac 
particle; so, it does tell us nothing about the actual 
(self-conjugate) character of $\psi_{\rm M}$, 
which can only be inferred from definition (\ref{2.1}).

	The current formal way of introducing 
a `self-conjugate' neutrino  is different from 
Majorana's way $-$ for a general review, see e.g. 
Ref.~$[{\ref{Bilenky1987}}]$ or $[{\ref{Giunti2007}}]$ $-$ 
and is essentially based on a `reformulation' of the 
Majorana neutrino theory in the light of parity-violating 
phenomenology $[{\ref{McLennan1957}},{\ref{Case1957}}]$. 
It leads to a neutrino type really compatible 
with the SM, being inspired by the idea of 
regarding a {\it massive} neutrino as merely 
an extension of a {\it massless} one. The new 
approach, in terms of basic {\it chiral} 
neutrino fields, relies just upon one peculiar
requirement naturally fulfilled by the original 
Majorana field $\psi_{\rm M}$: its being 
made up of two chiral components, 
$\frac{1}{2}(1 \mp \gamma^5)\psi_{\rm M}$, 
that are subjected to the {\it mutual link}
\begin{equation}
\frac{1}{2}(1 \pm \gamma^5)\psi_{\rm M} =
U_C \left[\frac{1}{2}(1 \mp \gamma^5)\psi_{\rm M}\right]^{\dagger {\rm T}}
\label{2.2ter}
\end{equation}

\noindent 
as a result of the condition $\psi_{\rm M}=\psi_{\rm M}^c$
(recall that $U_C \gamma^{5\dagger {\rm T}}=-\gamma^5 U_C$ 
and $U_C^{\dagger {\rm T}} = U_C^{-1}$). For a 
better insight into this point, let us start off by 
considering the purely left-handed (i.e. negative-chirality) 
neutrinos and purely right-handed (i.e. positive-chirality) 
antineutrinos as known from experience. As far as they 
are assumed to be massless, the question of the existence 
of opposite-chirality `complements' of their own fields 
may somehow be ignored. This, strictly speaking, 
can no longer be the case in the presence of 
(not exactly zero) neutrino masses. If one is 
in particular thinking of real neutrinos and 
antineutrinos as {\it standard} massive elementary 
fermions and antifermions, one should be able to supply 
their (originally massless) Lagrangians $-$ by adding 
e.g. suitable Higgs couplings $-$ with mass terms 
proportional to
\begin{equation}
{\bar {\psi}_{\rm L}}\psi_{\rm R}
+ {\bar {\psi}_{\rm R}}\psi_{\rm L}
= {\bar \psi}\psi
\label{2.2}
\end{equation}

\noindent
and
\begin{equation}
{\bar {\psi}^c_{\rm L}}\psi^c_{\rm R} 
+ {\bar {\psi}^c_{\rm R}}\psi^c_{\rm L}
= {\bar {\psi}^c}\psi^c,
\label{2.3}
\end{equation}

\noindent
respectively (${\bar {\psi}_{\rm L}}=
\psi^\dagger_{\rm L}\gamma^0$, and so on), 
where
\begin{equation}
\psi_{\rm L} \equiv \frac{1}{2}(1 - \gamma^5)\psi, \,\,\,
\psi_{\rm R} \equiv \frac{1}{2}(1 + \gamma^5)\psi
\label{2.4}
\end{equation}

\noindent
and
\begin{equation}
\psi^c_{\rm L} \equiv \frac{1}{2}(1 - \gamma^5)\psi^c,
\,\,\,
\psi^c_{\rm R} \equiv \frac{1}{2}(1 + \gamma^5)\psi^c. 
\label{2.5}
\end{equation}

\noindent
For clarity's sake, it is worth noting that here 
symbols $\psi_{\rm L,R}$ and $\psi^c_{\rm L,R}$ 
are used {\it quite symmetrically} to denote the 
left- and right-handed chiral components of $\psi$ 
and those of $\psi^c$; so, one in particular has 
$\psi^c_{\rm L,R} = (\psi^c)_{\rm L,R}$\,, 
and {\it not} $\psi^c_{\rm L,R} = (\psi^c)_{\rm R,L}$
(as often encountered in the literature). There appears 
to be, on the other hand, an alternative formal way 
of constructing a congruous mass term for a neutrino; 
it consists in {\it directly mixing} the two (left-handed) 
neutrino and (right-handed) antineutrino fields themselves 
$[{\ref{Jehle1949},\ref{Serpe1949}}]$. Doing like 
this, one is enabled to get, for instance, 
a scalar of the type
\begin{equation}
{\bar {\psi}_{\rm L}}\psi^c_{\rm R} + 
{\bar {\psi}^c_{\rm R}}\psi_{\rm L}
\equiv {\bar {\psi}'}\psi',
\label{2.6}
\end{equation}

\noindent
with the new (wholly `active') field 
\begin{equation}
\psi'(x) = \psi_{\rm L}(x) + \psi^c_{\rm R}(x)
\label{2.25}
\end{equation}

\noindent
replacing the original field $\psi(x) = 
\psi_{\rm L}(x) + \psi_{\rm R}(x)$. 
Note that (\ref{2.6}) is different 
from (\ref{2.2}) or (\ref{2.3}) {\it provided} 
$\psi^c_{\rm R} \not= \psi_{\rm R}\,,\,
\psi_{\rm L} \not=\psi^c_{\rm L}$. 
Such an approach to a massive neutrino does not 
necessarily need a complementary right-handed 
neutrino field, which could only enter into 
another (independent) mass term proportional to
\begin{equation}
{\bar {\psi}_{\rm R}}\psi^c_{\rm L} 
+ {\bar {\psi}^c}_{\rm L}\psi_{\rm R}.
\label{2.7} 
\end{equation}

\noindent
Of course, it is worth similarly stressing 
that this {\it extra} (wholly `sterile') 
massive neutrino type could by no means be 
conjectured if it were $\psi^c_{\rm L}
=\psi_{\rm L}\,,\,\psi_{\rm R}=\psi^c_{\rm R}$.
From (\ref{2.1ter}), recalling (\ref{2.4}) 
and (\ref{2.5}), it is therefore evident,
in particular, that a nontrivial definition of 
the new field variable given by (\ref{2.25}) 
(such that $\psi^c_{\rm R} \not= \psi_{\rm R},\,
\psi_{\rm L}\not=\psi^c_{\rm L}$) 
does indeed {\it prevent} it from being 
formally mistaken just for a $\psi_{\rm M}$ 
field variable. Yet, since
\begin{equation}
U_C \psi^{\dagger {\rm T}}_{\rm L}(x) = \psi^c_{\rm R}(x),
\;\;\;
U_C \psi^{c\dagger {\rm T}}_{\rm R}(x) = \psi_{\rm L}(x),
\label{2.7bis}
\end{equation}

\noindent 
one actually gets, in full analogy with (\ref{2.2ter}),
\begin{equation}
\psi'_{\rm R}(x) = U_C \psi'^{\dagger {\rm T}}_{\rm L}(x),
\;\;\;
\psi'_{\rm L}(x) = U_C \psi'^{\dagger {\rm T}}_{\rm R}(x),
\label{2.8}
\end{equation}

\noindent
with $\psi'_{\rm R}(x) = \frac{1}{2}(1 + \gamma^5)\psi'(x)
= \psi^c_{\rm R}(x)$ and $\psi'_{\rm L}(x)=\frac{1}{2}
(1 - \gamma^5)\psi'(x) = \psi_{\rm L}(x)$.
Hence it is commonly argued that Eq.~(\ref{2.25}) 
is just defining a {\it self-conjugate} neutrino field 
variable, associated with a new kind of mass term $-$ 
proportional to (\ref{2.6}) $-$ which is conventionally 
known as a {\it `Majorana mass' term}. Such a conclusion 
relies on the fact that, by use of either (\ref{2.7bis}) 
or (\ref{2.8}), one globally obtains
\begin{equation}
\psi'(x) = U_C \psi'^{\dagger {\rm T}}(x).
\label{2.8ter}
\end{equation}
	
	Of course, in interpreting (\ref{2.8ter}) as really a 
{\it sufficient} (besides necessary) condition to state that 
$\psi'(x)$ is self-conjugate, one is implicitly taking 
for granted that $U_C \psi'^{\dagger {\rm T}}(x)$ 
{\it does correspond} to the `charge conjugate' of $\psi'(x)$, 
or that $U_C \psi^{\dagger {\rm T}}_{\rm L}(x)=U_C 
\psi'^{\dagger {\rm T}}_{\rm L}(x)$ {\it does correspond}
to the `charge conjugate' of $\psi_{\rm L}(x)=\psi'_{\rm L}(x)$. 
If so, it should then result
\begin{equation}
\psi'^c(x) \equiv C \psi'(x) C^{-1}= U_C \psi'^{\dagger {\rm T}}(x),
\label{2.9'bis}
\end{equation}

\noindent
with the Majorana condition $\psi'(x)=\psi'^c(x)$ being 
automatically fulfilled. At first sight, since Eq.~(\ref{2.9'bis}) 
is the exact analogue of Eq.~(\ref{2.9bis}), there seem to be 
no reasons for questioning it. Despite this, consider e.g. 
the two Weyl equations, into which the Dirac equation can be 
split up on going over to the zero-mass limit. As is well-known 
$[{\ref{Esposito2013}}]$, the solutions of one Weyl equation 
amount to the couple of (massless) Dirac solutions
\begin{equation}
\psi_{\rm L}(x)\,,\,\psi_{\rm R}^c(x) 
= U_C \psi^{\dagger {\rm T}}_{\rm L}(x),
\label{2.10}
\end{equation}

\noindent
whereas those of the other Weyl equation amount to
the remaining couple of (massless) Dirac solutions
\begin{equation}
\psi_{\rm R}(x)\,,\,\psi_{\rm L}^c(x)
= U_C \psi^{\dagger {\rm T}}_{\rm R}(x).
\label{2.11}
\end{equation}

\noindent
Referring merely to the positive-energy contributions 
in the field expansions, we may in particular 
associate with the left-handed solutions, $\psi_{\rm L}$ 
and $\psi_{\rm L}^c$, a {\it negative helicity} 
and with the right-handed ones, $\psi_{\rm R}$ 
and $\psi_{\rm R}^c$, a {\it positive helicity}. 
Parity violation clearly occurs whenever (\ref{2.10}) 
and (\ref{2.11}) enter {\it asymmetrically}, 
and it becomes maximal just when only {\it one} 
couple of solutions is really involved. In the 
same way, as $C$ does {\it not} change helicities, 
also $C$ violation is expected to occur, with a 
maximal degree just when, for example, the only 
couple (\ref{2.10}) appears to be available. 
Nevertheless, if we rewrite (\ref{2.10}) as
\begin{equation}
\psi_{\rm L}(x)\,,\,\psi_{\rm R}^c(x) \equiv C\psi_{\rm L}(x)C^{-1},
\label{2.10bis}
\end{equation}

\noindent
and (\ref{2.11}), accordingly, as
\begin{equation}
\psi_{\rm R}(x)\,,\,\psi_{\rm L}^c(x) \equiv C\psi_{\rm R}(x)C^{-1},
\label{2.11bis}
\end{equation}

\noindent
then, even on admitting that solutions (\ref{2.11bis})
are suppressed, we {\it cannot} truly say that $C$ 
is violated, and we are in fact faced with a 
{\it helicity inverting} charge-conjugation operation! 
On the contrary, if we choose to interpret (\ref{2.10}) as
\begin{equation}
\psi_{\rm L}(x)\,,\,\psi_{\rm R}^c(x) \equiv C\psi_{\rm R}(x)C^{-1},
\label{2.10ter}
\end{equation}

\noindent
and (\ref{2.11}), accordingly, as
\begin{equation}
\psi_{\rm R}(x)\,,\,\psi_{\rm L}^c(x) \equiv C\psi_{\rm L}(x)C^{-1},
\label{2.11ter}
\end{equation}

\noindent
we immediately see that any asymmetry occurring
between (\ref{2.10ter}) and (\ref{2.11ter}) 
does {\it really} imply $C$ violation, with 
$C$ now being so defined as to {\it really} 
leave helicity unchanged. 

	In the light of these remarks on how to 
get an appropriate $C$ definition (not affecting 
helicity) for zero-mass spin-$\frac{1}{2}$ particles, 
it appears quite reasonable to try to check more 
carefully whether Eq.~(\ref{2.9'bis}) is really 
consistent or not in the general framework of 
standard QFT. For this purpose, in order to form 
a clear idea on how to proceed, it may be useful 
to look first at Eq.~(\ref{2.9bis}). As is well-known, 
primarily putting $\psi^c(x) \equiv C \psi(x) C^{-1}$ 
does indeed mean making reference to the {\it basic} 
definition of charge conjugation $C$, just coinciding 
with the {\it fundamental representation} of $C$ 
(in the fermion--antifermion Fock space). Take e.g. 
the standard normal mode expansion of $\psi(x)$ 
in terms of single `particle' annihilation operators
$a^{(h)}(\bf p)$ and `antiparticle' creation 
operators $b^{\dagger(h)}(\bf p)$ obeying the 
usual anticommutation rules and being relevant 
to simultaneous eigenstates of momentum $\bf p$ 
and helicity $h$. It looks like 
\begin{equation}
\psi(x) = \int \frac{d^3{\bf p}}{(2\pi)^3 2p^0} \sum_{h} \left[
a^{(h)}({\bf p}) u^{(h)}(p) e^{-ip\cdot x}
+ b^{\dagger(h)}({\bf p}) v^{(h)}(p) e^{ip\cdot x}
\right] 
\label{2.27}
\end{equation}

\noindent
($\hbar=c=1$), where $u^{(h)}(p)$ and $v^{(h)}(p)$ are
four-spinor coefficients depending on four-momentum $p \equiv 
(p^0,\bf p)$ $(p^0>0)$. We thus have, according to 
standard QFT, that $C \psi(x) C^{-1}$ is the net field 
obtained from (\ref{2.27}) {\it as a result of the 
transformation} $[{\ref{Merzbacher1970}}]$
\begin{equation}
C a^{(h)}({\bf p}) C^{-1} = b^{(h)}({\bf p}),
\;\;\;
C b^{\dagger(h)}({\bf p}) C^{-1} = a^{\dagger(h)}({\bf p}). 
\label{2.28} 
\end{equation}

\noindent
We also know that the subsequent equality, $C\psi(x)C^{-1}
=U_C\psi^{\dagger {\rm T}}(x)$, does instead tell us how 
the field $\psi^c(x) \equiv C \psi(x) C^{-1}$ as defined 
by means of (\ref{2.28}) can {\it equivalently} be obtained 
via a suitable mapping, $\psi(x) \longrightarrow \psi^c(x) = 
U_C\psi^{\dagger {\rm T}}(x)$, in the four-spinor space 
(such a mapping is indeed allowed by the fact that $\psi^{\dagger {\rm T}}$
and $\psi^c$ share all four freedom degrees corresponding 
to the actual `particle' and `antiparticle' helicity eigenstates). 
Of course, Eq.~(\ref{2.9bis}) applies as well 
to a field, $\psi_{\rm M}(x)$, having the manifestly 
self-conjugate form (\ref{2.1}), whose expansion can be 
obtained from (\ref{2.27}) by making the substitutions 
$a^{(h)}({\bf p}) \rightarrow a^{(h)}_{\rm M}({\bf p}) 
\;,\;b^{\dagger(h)}({\bf p}) \rightarrow a^{\dagger(h)}_{\rm M}
({\bf p})$, where $a^{(h)}_{\rm M}({\bf p}) = 
\frac{1}{\sqrt2}\left[a^{(h)}({\bf p}) + b^{(h)}({\bf p})\right]$ 
and $a^{\dagger(h)}_{\rm M}({\bf p}) = \frac{1}{\sqrt2}
\left[b^{\dagger(h)}({\bf p}) + a^{\dagger(h)}({\bf p})\right]$. 
Let us pass now to analyse Eq.~(\ref{2.9'bis}), addressed 
to a field, $\psi'(x)$, being such that
\begin{equation}
\psi'(x) = \frac{1}{2}(1 - \gamma^5)\psi(x) +
\frac{1}{2}(1 + \gamma^5)\psi^c(x),
\label{2.25bis}
\end{equation}

\noindent
with $\psi(x)\not=\psi^c(x)$. We shall have, 
quite similarly, that writing $\psi'^c(x) 
\equiv C \psi'(x) C^{-1}$ does mean 
{\it defining} $\psi'^c(x)$ {\it as just that field 
which is obtained from} $\psi'(x)$ {\it by merely 
turning every annihilation and creation operator 
into their respective charge-conjugates}. 
As such operators can only be found within the 
$\psi(x)$ and $\psi^c(x)$ expansions, we may 
also write, due to the {\it linearity} property 
of $C$ in the Fock space, 
\begin{equation}
\psi'^c(x) \equiv C \psi'(x) C^{-1} = 
\frac{1}{2}(1 - \gamma^5)C \psi(x) C^{-1} 
+ \frac{1}{2}(1 + \gamma^5) C \psi^c(x) C^{-1}.
\label{2.25ter}
\end{equation}

\noindent 
The actual check to be given to Eq.~(\ref{2.9'bis}) 
should therefore concern $C \psi'(x) C^{-1}
= U_C \psi'^{\dagger {\rm T}}(x)$. In other words: 
Is the mapping $\psi'(x) \longrightarrow \psi'^c(x) 
= U_C \psi'^{\dagger {\rm T}}(x)$ consistently 
providing, as claimed, an {\it equivalent} way to get 
$\psi'^c(x) \equiv C \psi'(x) C^{-1}$ from $\psi'(x)$? 
To answer this question, one has to do nothing else 
than {\it directly applying prescription} (\ref{2.25ter}). 
Doing like this, one obtains
\begin{equation}
\psi'^c(x) =
\frac{1}{2} (1 - \gamma^5) \psi^c(x)
+ \frac{1}{2} (1 + \gamma^5) \psi(x),
\label{2.12bis}
\end{equation}

\noindent
and hence, by use again of the compact notations 
(\ref{2.4}) and (\ref{2.5}), 
\begin{equation}
\psi'^c(x) = \psi^c_{\rm L}(x) + \psi_{\rm R}(x)
\not= \psi'(x) = \psi_{\rm L}(x) + \psi^c_{\rm R}(x).
\label{2.12} 
\end{equation}

\noindent
The key to (\ref{2.12}) is just the {\it linear} 
behavior of the fundamental representation of $C$, 
as particularly regards its action inside 
the {\it single} Dirac-field chiral components 
$\psi_{\rm L}$ and $\psi^c_{\rm R}$. Such 
an outcome, if carefully examined, should 
not seem so surprising, especially 
in view of what we already know from the 
$V-A$ theory $[{\ref{Feynman1958}-\ref{Sakurai1958}}]$. 
Looking at Eq.~(\ref{2.25bis}), we can easily 
realize that purely replacing every annihilation 
and creation operator with their own charge-conjugates 
does amount to globally making the {\it net} 
interchange $\psi \rightleftharpoons \psi^c$
relative to $\frac{1}{2}(1 \mp \gamma^5)$; 
and this, indeed, fully agrees with the general 
fact that, if one takes ${\bar \psi}_2\gamma^\mu\psi_1$ 
and ${\bar \psi}_2\gamma^\mu(-\gamma^5)\psi_1$ 
as Dirac-field bilinear {\it covariants} 
(as in the $V-A$ theory), then one gets 
${\bar \psi}_2\gamma^\mu(1 - \gamma^5)
\psi_1 \stackrel {C}{\rightarrow}{\bar {\psi}^c}_2
\gamma^\mu(1 - \gamma^5)\psi^c_1$ 
$[\ref{Sakurai1964}]$. A comparison 
of Eqs.~(\ref{2.8ter}) and (\ref{2.12}) 
gives, despite appearances, 
\begin{equation}
C \psi'(x) C^{-1} 
\not= U_C \psi'^{\dagger {\rm T}}(x),
\label{2.29}
\end{equation}

\noindent
and then
\begin{equation}
C \psi_{\rm L}(x) C^{-1} 
\not= U_C \psi^{\dagger {\rm T}}_{\rm L}(x);
\label{2.12ter}
\end{equation}

\noindent
which ultimately means that standard QFT 
does {\it not} really allow  Eq.~(\ref{2.9bis}) 
to be extended to $\psi'(x)$. By the way, the 
{\it explicit} `new' formal representations 
for $C \psi_{\rm L}(x) C^{-1}$ and $C \psi'(x) C^{-1}$
will be discussed in Secs.~3 and 4: see 
Eqs.~(\ref{2.16}) and (\ref{3.10}), respectively.  
It is obvious that, to avoid the inequalities (\ref{2.12}), 
(\ref{2.29}), and (\ref{2.12ter}), one could 
just assume $\psi=\psi^c=\psi'=\psi'^c$, 
but this would also {\it cancel} the 
distinction between an `active' ($\psi'$) and 
a `sterile' ($\psi'^c$) fermion field, as well 
as the distinction itself between a `Majorana' 
and a `Dirac' mass term! So, if the primary 
$C$-definition (\ref{2.25ter}) is truly 
relied upon, one is indeed led to conclude that 
{\it an `active' fermion field of the type} 
$\psi_{\rm L}(x) + (\psi^c)_{\rm R}(x)$ {\it and 
its `sterile' counterpart} $(\psi^c)_{\rm L}(x) 
+ \psi_{\rm R}(x)$ {\it are themselves a pair 
of mutually `charge conjugate' (rather than 
individually self-conjugate) fields}. 
It will be shown in Sec.~6 that the apparent 
(conventional) {\it full} neutrality of such fields 
is to be actually interpreted as a mere `neutrality' 
restricted to {\it scalar-type charges}. With 
the help of (\ref{2.12}), it can be checked that
\begin{equation}
\psi'(x) + \psi'^c(x) = \psi(x) + \psi^c(x).
\label{2.33}
\end{equation}

\noindent
This formally enables one to define the genuine 
Majorana field (\ref{2.1}) {\it also as} 
\begin{equation}
\psi_{\rm M}(x) = \frac{1}{\sqrt{2}}[\psi'(x) + \psi'^c(x)],
\label{2.34}
\end{equation}

\noindent
where $\psi'$ and $\psi'^c$ are exactly identical 
to the field components within square brackets 
in (\ref{2.1ter}). 

	The point at issue can also be faced 
in a reversed manner. We may begin, instead, by 
{\it assuming} field $\psi'$ to be truly self-conjugate, 
so that we are allowed to put $\psi'=\psi'^c$ 
($\propto \psi_{\rm M}$). Since either $\psi'$ 
or $\psi'^c$ is to be still meant as in 
Eq.~(\ref{2.12}), we shall then have as well
$\psi=\psi^c$ and, after all, $\psi'=\psi$.  
This automatically eliminates the inequalities 
in Eqs.~(\ref{2.12}) and (\ref{2.29}), but the 
chargeless fermion model obtained is {\it not} 
the same as the `Majorana mass' conventional one. 
First, it is evident that setting $\psi'=\psi'^c$ 
(in place of $\psi'\not=\psi'^c$) does cause fields 
$\psi'$ and $\psi'^c$ to {\it lose} their original 
`active' and `sterile' distinctive characters. 
Hence, in line with Eq.~(\ref{2.1ter}) or Eq.~(\ref{2.34}),
but {\it not} in line with what is usually 
claimed, the general conclusion is to be drawn that 
{\it there cannot exist two different $-$ `active' 
and `sterile' $-$ kinds of genuine (truly 
self-conjugate) Majorana fermions}. Moreover, 
we have to consider that the net equality $\psi'=\psi$ 
does unavoidably make a `Majorana mass' term 
(proportional to ${\bar {\psi}'}\psi'$) 
{\it indistinguishable} from a `Dirac mass' 
term (proportional to ${\bar {\psi}}\psi$). 
So, it must be concluded as well that 
{\it a stricly neutral spin-$\frac{1}{2}$ 
fermion which is supposed to bear a `Majorana mass' 
can likewise be said to bear a `Dirac mass' 
(or vice versa)}. In other words, there would be 
simply a {\it unified} mass kind for such a fermion, 
which may be equally taken for a `Majorana' as for 
a `Dirac' mass kind\,! This fully corresponds to 
the fact that we cannot manage to build more than 
{\it one} mass term from chiral fields like those 
in (\ref{2.2bis}).

\section{On the truly orthodox way of applying  
particle--antiparticle conjugation to single 
chiral fields within standard QFT}	
	Owing to its subtlety, the whole question 
raised above deserves an even more detailed analysis. 
As shown by (\ref{2.12ter}), and as already implied 
in the opening discussion on Weyl solutions, 
the core of the problem is just how to define 
the `charge conjugates' of the {\it unpaired} 
chiral projections, $\psi_{\rm L}$ and $\psi^c_{\rm R}$, 
entering into (\ref{2.25}). To start with, it is 
worth emphasizing that the standard Dirac-field 
`prescription' (\ref{2.9bis}) does {\it not} 
automatically extend to the individual chiral 
components of $\psi$. In principle, we may write
\begin{equation}
U_ C\psi^{\dagger {\rm T}} 
= U_C \left[\frac{1}{2}(1 - \gamma^5)\psi\right]^{\dagger {\rm T}}
+ U_C \left[\frac{1}{2}(1 + \gamma^5)\psi\right]^{\dagger {\rm T}}
\label{2.18}
\end{equation}

\noindent
{\it as well as}
\begin{equation}
U_C\psi^{\dagger {\rm T}} = \frac{1}{2}(1 - \gamma^5)U_C\psi^{\dagger {\rm T}}
+ \frac{1}{2}(1 + \gamma^5)U_C\psi^{\dagger {\rm T}},
\label{2.18bis}
\end{equation}

\noindent
where $\frac{1}{2}(1 \mp \gamma^5)U_C\psi^{\dagger {\rm T}} =
U_C\left[\frac{1}{2}(1 \pm \gamma^{5})\psi\right]^{\dagger {\rm T}}$. 
From a strict formal viewpoint, we are thus faced 
with two possible alternative ways of defining the 
`charge conjugates' of $\psi_{\rm L}$ and $\psi_{\rm R}$: 
we may put {\it either}  
\begin{equation}
\left\{ \!
\begin{array}{lcr}
C \psi_{\rm L}(x) C^{-1} \equiv 
U_C\psi_{\rm L}^{\dagger {\rm T}}(x) = (\psi^c)_{\rm R}(x)
\\[.1in]  
C \psi_{\rm R}(x) C^{-1} \equiv 
U_C\psi_{\rm R}^{\dagger {\rm T}}(x) = (\psi^c)_{\rm L}(x)
\end{array}
\right.
\label{2.21}
\end{equation}

\noindent
{\it or} 
\begin{equation}
\left\{ \!
\begin{array}{lcr}
C \psi_{\rm L}(x) C^{-1} \equiv
U_C\psi_{\rm R}^{\dagger {\rm T}}(x) = (\psi^c)_{\rm L}(x)
\\[.1in]
C \psi_{\rm R}(x) C^{-1} \equiv 
U_C\psi_{\rm L}^{\dagger {\rm T}}(x) = (\psi^c)_{\rm R}(x)
\end{array}
\right.
\label{2.22}
\end{equation}

\noindent 
and both these assumptions lead to {\it the same 
overall result}
\begin{equation}
\psi^c = (\psi_{\rm L} + \psi_{\rm R})^c 
= U_C (\psi^{\dagger {\rm T}}_{\rm L} + \psi^{\dagger {\rm T}}_{\rm R}) 
= U_C (\psi^{\dagger {\rm T}}_{\rm R} + \psi^{\dagger {\rm T}}_{\rm L})
= U_C \psi^{\dagger {\rm T}}.
\label{2.23}
\end{equation}

\noindent
Actually, that there may be some `freedom' in defining 
a $C$ operation is not a novelty in the literature: 
see e.g. the quite similar reasonings made in 
Ref.~$[{\ref{Ziino2006}}]$, or those made 
in  Ref.~$[{\ref{Dvoeglazov1997}}]$ by use of the 
`chiral helicity' special construct, and see as well 
Ref.~$[{\ref{Dvornikov2012}}]$, where a new interpretation 
of the Majorana condition is proposed. In comparison with 
Refs.~$[{\ref{Dvoeglazov1997}}]$ and $[{\ref{Dvornikov2012}}]$, 
the major distinctive feature can here be drawn from Eq.~(\ref{2.23}), 
which shows that the helicity of a massive spin-$\frac{1}{2}$ 
fermion is always preserved by $C$ no matter whether (\ref{2.21}) 
or (\ref{2.22}) is being adopted. Also, note that both 
in (\ref{2.21}) and in (\ref{2.22}) the matrix $U_C$ is 
regularly connecting field components with {\it opposite} 
chiralities. It is worth remarking, however, that in 
the zero-mass limit a `$C$ definition' like (\ref{2.21}) 
leads to the couples of solutions (\ref{2.10bis}) and 
(\ref{2.11bis}), whereas a `$C$ definition' like (\ref{2.22}) 
leads to the {\it alternative} couples of  solutions 
(\ref{2.10ter}) and (\ref{2.11ter}). To tell which 
of the two options (\ref{2.21}) and (\ref{2.22}) 
is to be taken as the {\it truly orthodox} one according 
to standard QFT, it is enough to consider that only 
(\ref{2.22}) is strictly consistent with the {\it genuine} 
(primary) $C$ definition (\ref{2.28}). The remaining option, 
even though it is just as well allowed by (\ref{2.23}) 
and may all the same reproduce Eq.~(\ref{2.9bis}), 
would rather correspond to a $C$ operation whose 
fundamental representation does no longer appear to be 
{\it fully} defined inside the (conventional) Fock space, 
as it would now consist of (\ref{2.28}) supplemented 
by one `spurious' basic prescription, $\frac{1}{2}
(1 \mp \gamma^5) \rightarrow \frac{1}{2}(1 \pm \gamma^5)$, 
{\it outside} the Fock space. To this it should be added 
that the choice of either (\ref{2.21}) or (\ref{2.22}) 
$-$ apparently irrelevant if we neglect (\ref{2.28}) 
and we restrict ourselves to Eq.~(\ref{2.9bis}) $-$ 
is made by no means irrelevant if the validity of 
Eq.~(\ref{2.9'bis}) is also invoked, with $\psi'(x)$ 
being defined as in (\ref{2.25}). The point is that 
Eq.~(\ref{2.9'bis}) may really be claimed to hold 
{\it only if} the `wrong' option (\ref{2.21}) $-$ 
just defining $\psi_{\rm L}$ and $(\psi^c)_{\rm R}$ 
as the `charge conjugates' of each other $-$ is adopted. 
So, after all, we may even choose (\ref{2.21}) 
(as usually done) to extend Eq.~(\ref{2.9bis}) 
to a field like $\psi'(x)$, but in this way we are 
not keeping to the {\it strict} QFT prescription 
(\ref{2.28}) and we are actually introducing a 
{\it new} `charge conjugation' operation, say, 
$C'$, which is such that $C' \psi'(x) C'^{-1}= 
U_C \psi'^{\dagger {\rm T}}$ and $C'\psi_{\rm L,R}(x)C'^{-1} 
= U_C\psi_{\rm L,R}^{\dagger {\rm T}}(x)$, 
and which should not be confused with the one, $C$ 
itself, rigorously acting as `particle--antiparticle 
conjugation' according to (\ref{2.28}). It will
be shown in Sec.~6 that $C'$ (normally mistaken 
for $C$) does amount, more precisely, to a mere 
`scalar-charge conjugation' operation. 

	The difference between two such ways 
of defining a `charge conjugation' operation 
is clearly brought to its extreme consequences 
(which affect helicities themselves) on passing 
to the zero-mass case. This has already been 
mentioned in the previous section; yet, for 
completeness' sake, it may be worth trying 
to deal with that case in more detail, too. 
The zero-mass limit leads to a special situation 
in which the matrix $U_C$ can `manifestly' be seen 
to connect the two chiralities $[{\ref{Itzykson1985}}]$. 
This is due to the well-known fact that 
positive- and negative-energy massless 
eigenspinors with discordant (concordant) 
helicities are themselves {\it chiral} spinors 
with concordant (discordant) chiralities. 
So, if $\psi_{\rm L}(x)$ and $\psi_{\rm R}(x)$ 
are now taken as zero-mass fields, their 
respective normal mode expansions will 
become
\begin{equation}
\psi_{\rm L}(x) = \int \frac{d^3{\bf p}}{(2\pi)^3 2p^0}  
\left[ a^{(-)}({\bf p}) u^{(-)}(p) e^{-ip\cdot x}
+ b^{\dagger(+)}({\bf p}) v^{(+)}(p) e^{ip\cdot x}
\right] 
\label{2.30}
\end{equation}

\noindent
and
\begin{equation}
\psi_{\rm R}(x) = \int \frac{d^3{\bf p}}{(2\pi)^3 2p^0}  
\left[ a^{(+)}({\bf p}) u^{(+)}(p) e^{-ip\cdot x}
+ b^{\dagger(-)}({\bf p}) v^{(-)}(p) e^{ip\cdot x}
\right], 
\label{2.31}
\end{equation}

\noindent
where the superscripts $(\mp)$ still denote 
the eigenvalues of the helicity variable $h$. 
Hence, recalling that particle--antiparticle 
conjugation $C$ leaves (by definition) helicity 
{\it unvaried} $[{\ref{Merzbacher1970}}]$, we find 
e.g. that the `charge conjugate' of the (massless) field 
$\psi_{\rm L}(x)$ is to be strictly defined as
\begin{equation}
C \psi_{\rm L}(x) C^{-1} \equiv
U_C \psi^{\dagger {\rm T}}_{\rm R}(x) = (\psi^c)_{\rm L}(x),
\label{2.13}
\end{equation}

\noindent
with $U_C\psi^{\dagger {\rm T}}_{\rm R}(x)$ including 
every required positive-energy spinor $U_Cv^{(-)}(p)$, 
of {\it negative} helicity, as well as every required 
negative-energy spinor $U_Cu^{(+)}(p)$, of {\it positive} 
helicity, and {\it not} as
\begin{equation}
C \psi_{\rm L}(x) C^{-1} \equiv
U_C \psi^{\dagger {\rm T}}_{\rm L}(x) = (\psi^c)_{\rm R}(x),
\label{2.14}
\end{equation}

\noindent
with $U_C\psi^{\dagger {\rm T}}_{\rm L}(x)$ containing 
the corresponding {\it unwanted} spinors having
interchanged helicities. On the other hand, 
this fully agrees with what has been inferred 
in Sec.~2 from comparing (\ref{2.10ter}),(\ref{2.11ter}) 
with (\ref{2.10bis}),(\ref{2.11bis}): it is 
definition (\ref{2.13}), and {\it not} definition 
(\ref{2.14}), that strictly requires also the presence 
of a (right-handed) field solution $\psi_{\rm R}$ 
(quite missing in neutrino phenomenology) 
thus leading us to conclude that $C$ itself 
is (maximally) {\it violated} by neutrino physics! 
Of course, we may come to (\ref{2.13}) even without 
allowing for the specific normal mode expansions 
(\ref{2.30}) and (\ref{2.31}): it is sufficient 
simply to take account of the basic $C$-definition 
(\ref{2.28}) to get 
\begin{equation}
C \psi_{\rm L}(x) C^{-1} = 
\frac{1}{2} (1 - \gamma^5) C \psi(x) C^{-1}
= \frac{1}{2} (1 - \gamma^5)U_C \psi^{\dagger {\rm T}}(x),
\label{2.15}
\end{equation}

\noindent
where the former equality (leaving chirality unaffected)
is just due to the {\it linear} behavior of $C$ in the
Fock space. Hence it also follows, as already implied 
in (\ref{2.13}),
\begin{equation}
C \psi_{\rm L}(x) C^{-1} =
U_C \left[\frac{1}{2} (1 + \gamma^5) \psi(x)\right]^{\dagger {\rm T}}
= U_C \psi_{\rm R}^{\dagger {\rm T}}(x),
\label{2.16}
\end{equation}

\noindent
and it may be concluded that {\it a $C$-matrix 
identical with the conventional one is still 
available, provided that complex conjugation 
is supplemented by} `${\rm L} \rightarrow {\rm R}$' 
{\it exchange}. This, of course, does not affect 
the `whole' Dirac result $\psi^c=U_C\psi^{\dagger {\rm T}}$ 
(with $\psi=\psi_{\rm L}+\psi_{\rm R}$) $-$ 
see Eq.(\ref{2.23}) $-$ and it is just the way 
to maintain helicity invariance under $C$. 
Note, on the other hand, that {\it no} $4\times4$ 
matrix $U'_C$ can further be found being such that 
$C \psi_{\rm L}(x) C^{-1}=U_C \psi_{\rm R}^{\dagger {\rm T}}(x)
=U'_C\psi_{\rm L}^{\dagger {\rm T}}(x)$: to realize it, 
one may simply consider that positive-energy 
antifermions (fermions) annihilated (created) 
by a massless field like $U'_C\psi_{\rm L}^{\dagger {\rm T}}(x)$, 
with
\begin{equation}
\psi_{\rm L}^{\dagger {\rm T}}(x) = \int \! \frac{d^3{\bf p}}{(2\pi)^3 2p^0}  
\left[ b^{(+)}({\bf p}) v^{(+) \dagger {\rm T}}(p) e^{-ip\cdot x}
+ a^{\dagger(-)}({\bf p}) u^{(-) \dagger {\rm T}}(p) e^{ip\cdot x}
\right], 
\label{2.24}
\end{equation}

\noindent
would rather have {\it inverted} helicities 
with respect to positive-energy fermions 
(antifermions) annihilated (created) by 
$\psi_{\rm L}(x)$. On the grounds of either 
Eq.~(\ref{2.13}) or Eq.~(\ref{2.15}), 
it can thus be argued, after all, that 
even for zero mass there is no {\it net} 
chirality flip actually induced by 
particle--antiparticle conjugation (\ref{2.28}) 
(despite the unquestionable fact that 
$U_C$ itself is manifestly connecting 
fields with {\it opposite} chiralities!).

\section{A double variety of mutually `charge 
conjugate' spin-1/2 fermion fields}
Take Eq.~(\ref{2.33}), and consider the two 
(both admissible) formal ways, either (\ref{2.1}) 
or (\ref{2.34}), of defining a (manifestly 
self-conjugate) Majorana field $\psi_{\rm M}(x)$ 
in terms of {\it mutually `charge conjugate'} 
spin-$\frac{1}{2}$ fermion fields, either $\psi(x),
\psi^c(x)$ or $\psi'(x),\psi'^c(x)$. These 
field pairs (no matter which of them is taken as
a mass-eigenfield pair) can be formally put 
on an {\it equal} footing via the {\it unitary 
transformation}  
\begin{equation}
\left\{
\begin{array}{lcl}
 \psi'(x)\!\!&=&\!\!X_{\rm L} \psi(x) + X_{\rm R} \psi^c(x) \\ [0.05in] 
 \psi'^c(x)\!\!&=&\!\!X_{\rm R} \psi(x) + X_{\rm L} \psi^c(x) 
\end{array}
\right.
\label{3.1}
\end{equation}

\noindent
or the {\it inverse} one
\begin{equation}
\left\{
\begin{array}{lcl}
 \psi(x)\!\!&=&\!\!X_{\rm L} \psi'(x) + X_{\rm R} \psi'^c(x) \\ [0.05in] 
 \psi^c(x)\!\!&=&\!\!X_{\rm R} \psi'(x) + X_{\rm L} \psi'^c(x) 
\end{array}
\right.
\label{3.2bis}
\end{equation}

\noindent
where
\begin{equation}
X_{\rm L} \equiv \frac{1}{2}(1 - \gamma^5),
\;\;\;\;
X_{\rm R} \equiv \frac{1}{2}(1 + \gamma^5)
\label{3.2}
\end{equation}

\noindent
(and where, of course, $X_{\rm L}X_{\rm R}=X_{\rm R}X_{\rm L}
=0,X_{\rm L,R}^\dagger=X_{\rm L,R},X_{\rm L,R}^2
=X_{\rm L,R},X_{\rm R} + X_{\rm L}=1,X_{\rm R} 
- X_{\rm L}=\gamma^5$).
 
	Transformations (\ref{3.1}) and (\ref{3.2bis}),
which can be suitably rewritten in the matrix forms
\begin{equation}
\pmatrix{\psi'(x)\cr \psi'^c(x)\cr} \!\!=\!\!
\pmatrix{X_{\rm L}&X_{\rm R}\cr X_{\rm R}&X_{\rm L}\cr}
\pmatrix{\psi(x)\cr \psi^c(x)\cr},
\pmatrix{\psi(x)\cr \psi^c(x)\cr} \!\!=\!\!
\pmatrix{X_{\rm L}&X_{\rm R}\cr X_{\rm R}&X_{\rm L}\cr}
\pmatrix{\psi'(x)\cr \psi'^c(x)\cr},
\label{3.2ter}
\end{equation}

\noindent
make it possible to introduce a generalized 
matrix representation for $C$. Such an extension, 
clearly superfluous on dealing merely with the 
field pair $\psi(x),\psi^c(x)$, turns out to be 
strictly needed if the field pair $\psi'(x),\psi'^c(x)$ 
is also allowed for. This is because the use 
of Eq.~(\ref{2.16}) $-$ along with $C \psi^c_{\rm R}(x) C^{-1} 
= U_C \psi_{\rm L}^{c \dagger {\rm T}}(x)$ $-$ 
can only lead to the {\it trivial} overall result
\begin{equation}
C \psi'(x) C^{-1} \equiv \psi'^c(x) = U_C \psi'^{c \dagger {\rm T}}(x),
\label{3.10}
\end{equation}

\noindent
where $U_C \psi'^{c \dagger {\rm T}}(x)$ 
is merely an identical way to write $\psi'^c(x)$, 
and {\it not} an actual prescription to obtain 
$\psi'^c(x)$ {\it from} $\psi'(x)$\,! So, unlike what 
happens for $\psi(x) \stackrel{C}{\rightarrow} \psi^c(x)$,
with $\psi^c(x)\equiv C\psi(x)C^{-1}=U_C\psi^{\dagger {\rm T}}(x)$, 
there seems to be {\it no} $4\times4$ matrix $U'_C$ 
effectively representing $\psi'(x) \stackrel{C}{\rightarrow}\psi'^c(x)$ 
and being such that $\psi'^c(x)\equiv C\psi'(x)C^{-1}=U'_C\psi'^{\dagger {\rm T}}(x)$
(the problem does not apply, of course, to a 
genuine Majorana field $\psi_{\rm M}$, since 
$\psi^c_{\rm M}=\psi_{\rm M}=U_C \psi^{c \dagger {\rm T}}_{\rm M}
=U_C \psi^{\dagger {\rm T}}_{\rm M}$). 
A deep motivation for this lack can be found when the 
normal mode expansions of $\psi'(x)$ and $\psi'^c(x)$ 
(taken as mass eigenfields) are derived $-$ 
see Eqs.~(\ref{2.35}) and (\ref{2.36}) below 
$-$ and it is realized that (quite differently 
from the `Dirac mass' case) there are indeed 
{\it neither any `particle' nor any `antiparticle' 
helicity freedom degrees shared by these expansions}. 
More precisely, one still has that the freedom degrees 
in question are {\it four} in all (as in the `Dirac mass' 
case) but one also has that only {\it two} of them 
are included in the $\psi'(x)$ expansion and the other 
{\it two} in the $\psi'^c(x)$ expansion. It thus follows 
that $\psi'(x)$, as given by Eq.~(\ref{2.35}), and 
$\psi'^c(x)$, as given by Eq.~(\ref{2.36}), are 
actually belonging to two {\it orthogonal} field spaces; 
and this clearly makes quite inadmissible any mutual 
link of the type $\psi'^c(x)=U'_C\psi'^{\dagger {\rm T}}(x)$
(which would instead require {\it one and the same} 
four-spinor field space for them both!). That said, 
let us begin by defining the `charge conjugate' 
of the column matrix 
\begin{equation}
\Psi(x) \equiv \pmatrix{\psi(x)\cr \psi^c(x)\cr}.
\label{3.3ter}
\end{equation}

\noindent 
We clearly have
\begin{equation}
C\Psi(x)C^{-1} \equiv \pmatrix{\psi^c(x)\cr \psi(x)\cr}
= \pmatrix{U_C&0\cr 0&U_C\cr}
\pmatrix{\psi^{\dagger {\rm T}}(x)\cr \psi^{c \dagger {\rm T}}(x)\cr},
\label{3.3}
\end{equation}

\noindent
where $\pmatrix{U_C&0\cr 0&U_C\cr}$ is an
$8\times8$ matrix (still made up, as expected, 
of $4\times4$ diagonal blocks). To obtain, 
likewise, the `charge conjugate' of 
\begin{equation}
\Psi'(x) \equiv \pmatrix{\psi'(x)\cr \psi'^c(x)\cr},
\label{3.3quater}
\end{equation}

\noindent
it must be borne in mind that $C$ fundamentally acts 
{\it just} on the annihilation and creation operators 
included in the fields, without affecting the 
transformation matrix in either (\ref{3.1}) or 
(\ref{3.2bis}). The result is
\begin{equation}
C\Psi'(x)C^{-1} \equiv \pmatrix{\psi'^c(x)\cr \psi'(x)\cr}
=\pmatrix{X_{\rm L}&X_{\rm R}\cr X_{\rm R}&X_{\rm L}\cr}
\pmatrix{\psi^c(x)\cr \psi(x)\cr}
\label{3.4}
\end{equation}

\noindent
or ultimately, in full agreement with (\ref{3.10}),
\begin{equation}
C\Psi'(x)C^{-1} \equiv \pmatrix{\psi'^c(x)\cr \psi'(x)\cr} 
= \pmatrix{0&U_C\cr U_C&0\cr}
\pmatrix{\psi'^{\dagger {\rm T}}(x)\cr \psi'^{c \dagger {\rm T}}(x)\cr},
\label{3.10bis}
\end{equation}

\noindent
where we have substituted both (\ref{3.3}) and
\begin{equation}
\pmatrix{\psi^{\dagger {\rm T}}(x)\cr \psi^{c \dagger {\rm T}}(x)\cr} = 
\pmatrix{X_{\rm L}^{\dagger {\rm T}}&X_{\rm R}^{\dagger {\rm T}}\cr 
X_{\rm R}^{\dagger {\rm T}}&X_{\rm L}^{\dagger {\rm T}}\cr}
\pmatrix{\psi'^{\dagger {\rm T}}(x)\cr \psi'^{c \dagger {\rm T}}(x)\cr}
\label{3.1bis*}
\end{equation}

\noindent
(and we have taken into account that 
$U_C X_{\rm L,R}^{\dagger {\rm T}} = X_{\rm R,L} U_C$).
It thus turns out that the {\it transformed} 
$8\times8$ matrix 
\begin{equation} 
\pmatrix{X_{\rm L}&X_{\rm R}\cr 
X_{\rm R}&X_{\rm L}\cr}
\pmatrix{U_C&0\cr 0&U_C\cr}
\pmatrix{X_{\rm L}^{\dagger {\rm T}}&X_{\rm R}^{\dagger {\rm T}}\cr
X_{\rm R}^{\dagger {\rm T}}&X_{\rm L}^{\dagger {\rm T}}\cr}
=\pmatrix{0&U_C\cr U_C&0\cr}
\label{3.10ter}
\end{equation}

\noindent
is {\it no more} trivially made up of diagonal blocks.
This generalized matrix representation of $C$, 
given by Eqs.~(\ref{3.3}) and (\ref{3.10bis}),
has been built {\it without necessarily specifying}
which of the two field pairs $\psi(x),\psi^c(x)$ 
and $\psi'(x),\psi'^c(x)$ should be also a mass-eigenfield 
pair: what has been only assumed is the validity 
of the formal link (\ref{2.9bis}) connecting $\psi(x)$ 
and $\psi^c(x)$. So, after all, Eqs.~(\ref{3.3}) and 
(\ref{3.10bis}) may be referred to as just the 
{\it basic} peculiar features generally distinguishing 
$\psi(x),\psi^c(x)$ and $\psi'(x),\psi'^c(x)$.
	
	By use of either (\ref{3.1}) or (\ref{3.2bis})
(and with the help of the $X_{\rm L,R}$ properties) 
it can further be checked that
\begin{equation}
\bar{\psi}\gamma^\mu\psi - \bar{\psi}^c
\gamma^\mu\psi^c =  \bar{\psi}'\gamma^\mu
(-\gamma^5)\psi' - \bar{\psi}'^c\gamma^\mu
(-\gamma^5)\psi'^c,
\label{4.1}
\end{equation}
  
\noindent
where the individual currents
\begin{equation}
\bar{\psi}'\gamma^\mu(-\gamma^5)\psi'
=  \bar{\psi}'\gamma^\mu X_{\rm L}\psi'
- \bar{\psi}'\gamma^\mu X_{\rm R}\psi'
=  \bar{\psi}\gamma^\mu X_{\rm L}\psi
- \bar{\psi}^c\gamma^\mu X_{\rm R}\psi^c
\label{4.2}
\end{equation}

\noindent
and  
\begin{equation}
\bar{\psi}'^c\gamma^\mu(-\gamma^5)\psi'^c
=  \bar{\psi}'^c\gamma^\mu X_{\rm L}\psi'^c
- \bar{\psi}'^c\gamma^\mu X_{\rm R}\psi'^c
=  \bar{\psi^c}\gamma^\mu X_{\rm L}\psi^c
- \bar{\psi}\gamma^\mu X_{\rm R}\psi
\label{4.2bis}
\end{equation}

\noindent
(just like the ordinary ones $\bar{\psi}\gamma^\mu\psi$ 
and $\bar{\psi}^c\gamma^\mu\psi^c$) are generally 
{\it non}vanishing [{\ref{Dvoeglazov2012}}], in spite 
of the fact that $\psi'(x) = U_C \psi'^{\dagger {\rm T}}(x)$ 
and $\psi'^c(x) = U_C \psi'^{c \dagger {\rm T}}(x)$.

	Let us look first at the 
`Dirac mass' special case, i.e. when, 
as usual, the field pair $\psi(x),\psi^c(x)$ 
is also a pair of mass eigenfields, 
the one defined by the expansion 
(\ref{2.27}) and the other by the 
charge conjugate of (\ref{2.27}). 
In this case, while $\psi(x)$ and 
$\psi^c(x)$ $-$ taken as free fields 
$-$ are {\it single} solutions 
of the Dirac equation (with a given 
mass parameter $m$), the same cannot 
be said for $\psi'(x)$ and $\psi'^c(x)$, 
as it will clearly result
\begin{equation}
i\gamma^\mu \partial_\mu \psi'(x) = m \psi'^c(x),
\;\;\;\;
i\gamma^\mu \partial_\mu \psi'^c(x) = m \psi'(x)
\label{4.3}
\end{equation}

\noindent
($\hbar=c=1$). Yet, on the basis of (\ref{4.1}), 
(\ref{4.2}), and (\ref{4.2bis}), it may be argued 
that fields $\psi'(x)$ and $\psi'^c(x)$ themselves 
seem in particular to enter into maximally-$P$-violating 
weak couplings like real `dynamical eigenfields' $-$ 
the former {\it wholly `active'} and the latter 
{\it wholly `sterile'} $-$ thus giving, furthermore, 
a {\it direct} evidence of the maximum $C$-violation 
also implied in those couplings. 

	Let us pass now to the  `Majorana mass' 
special case, i.e. when, on the contrary, it is 
just the field pair $\psi'(x),\psi'^c(x)$ 
that stands for an actual pair of mass 
eigenfields. To make an explicit derivation 
of the normal mode expansions defining 
$\psi'(x)$ and $\psi'^c(x)$ in such a case, 
we may exploit the fact that these expansions 
should clearly have forms which are also 
available for the zero-mass limit. Splitting up 
both $\psi'$ and $\psi'^c$ into chiral 
components, such that
\begin{equation}
\left\{ 
\begin{array}{lcr}
\frac{1}{2}(1 - \gamma^5)\psi' = \frac{1}{2}(1 - \gamma^5)\psi,
\;\;
\frac{1}{2}(1 + \gamma^5)\psi' = \frac{1}{2}(1 + \gamma^5)\psi^c
\\[.1in]  
\frac{1}{2}(1 - \gamma^5)\psi'^c = \frac{1}{2}(1 - \gamma^5)\psi^c,
\;\;
\frac{1}{2}(1 + \gamma^5)\psi'^c = \frac{1}{2}(1 + \gamma^5)\psi,
\end{array}
\right.
\label{2.37}
\end{equation}

\noindent
we may, thus, simply substitute the zero-mass normal 
mode expansions of the single chiral-field couples 
$\frac{1}{2}(1-\gamma^5)\psi,\frac{1}{2}(1+\gamma^5)\psi^c$ 
and $\frac{1}{2}(1-\gamma^5)\psi^c,\frac{1}{2}(1+\gamma^5)\psi$ 
to obtain (for the {\it non}zero-mass case at issue)
\begin{eqnarray}
\psi'(x) & \!\!=\!\! & \int \frac{d^3{\bf p}}{(2\pi)^3 2p^0}  
\left\{\big[a^{(-)}({\bf p}) u^{(-)}(p) + 
b^{(+)}({\bf p}) u^{(+)}(p)\big] 
e^{-ip\cdot x}\right.  \nonumber \\
& \! \!  & \left. + \big[b^{\dagger(+)}({\bf p}) v^{(+)}(p) 
+ a^{\dagger(-)}({\bf p}) v^{(-)}(p)\big] e^{ip\cdot x}
\right\}
\; \mbox{(`Major. mass')}
\label{2.35}
\end{eqnarray}

\noindent
and
\begin{eqnarray}
\psi'^c(x) & \!\!\!=\!\!\! & \int \frac{d^3{\bf p}}{(2\pi)^3 2p^0}  
\left\{\big[b^{(-)}({\bf p}) u^{(-)}(p) + 
a^{(+)}({\bf p}) u^{(+)}(p)\big] 
e^{-ip\cdot x}\right.  \nonumber \\
& \! \! & \left. + \big[a^{\dagger(+)}({\bf p}) v^{(+)}(p) 
+ b^{\dagger(-)}({\bf p}) v^{(-)}(p)\big] e^{ip\cdot x}
\right\}
\; \mbox{(`Major. mass')},
\label{2.36}
\end{eqnarray}

\noindent
with the superscripts $(\mp)$ still denoting 
(negative and positive) helicities. Of course, 
as now the other two fields, $\psi(x)$ and 
$\psi^c(x)$, are no longer mass eigenfields, 
it is not surprising that their corresponding 
expansions obtained from (\ref{2.35}) and
(\ref{2.36}) by use of (\ref{3.2bis}) may 
not be the same as the (usual) ones for the 
`Dirac mass' case. A glance at (\ref{2.35}) and 
(\ref{2.36}) shows that only operators such as 
$a^{(-)}({\bf p})$ and $b^{\dagger(+)}({\bf p})$ 
(plus their adjoints) enter into (\ref{2.35}), 
and similarly, only operators such as 
$b^{(-)}({\bf p})$ and $a^{\dagger(+)}({\bf p})$ 
(plus their adjoints) enter into (\ref{2.36}). 
This, at first sight, might lead one to mistake 
either $\psi'(x)$ or $\psi'^c(x)$ {\it alone} 
for a truly neutral spin-$\frac{1}{2}$ field, 
merely endowed with {\it two} freedom degrees 
like $a'^{(-)}({\bf p})\equiv a^{(-)}({\bf p})\,, 
\,a'^{(+)}({\bf p})\equiv b^{(+)}({\bf p})$ 
(as regards $\psi'$) or $b'^{(-)}({\bf p})\equiv b^{(-)}({\bf p})
\,, \,b'^{(+)}({\bf p})\equiv a^{(+)}({\bf p})$ 
(as regards $\psi'^c$). From comparing (\ref{2.35}) 
and (\ref{2.36}), it appears evident, however, 
that such pairs of annihilation operators, 
as well as the whole fields $\psi'(x)$ and $\psi'^c(x)$ 
themselves, are actually {\it interchanged} by 
particle--antiparticle conjugation (\ref{2.28}), 
despite the fact that (\ref{2.35}) and (\ref{2.36}) 
also provide clear evidence for the 
{\it individual} formal constraints 
$\psi'(x)=U_C\psi'^{\dagger {\rm T}}(x)$ 
and $\psi'^c(x)=U_C\psi'^{c \dagger {\rm T}}(x)$.
To this it is worth adding that (as it should 
be expected) the two expansions (\ref{2.35}) 
and (\ref{2.36}) are fully consistent with 
Eq.~(\ref{3.10bis}). Therefore, to come to a 
{\it genuine} self-conjugate field, it would 
be strictly necessary to impose the {\it extra} 
constraint $a^{(\mp)}({\bf p})=b^{(\mp)}({\bf p}), 
b^{\dagger(\pm)}({\bf p})=a^{\dagger(\pm)}({\bf p})$; 
which, indeed, would mean nothing else than {\it trivially 
reducing $\psi'$ and $\psi'^c$ to one and the same 
field by means of the Majorana condition $\psi'=\psi'^c$}\,! 
On the other hand, if we truly admit that (\ref{2.35}) 
and (\ref{2.36}) are just defining `charged' fields,
we have also to admit, of course, that standard 
(i.e. scalar-type) charges are to be ruled out for them. 
It is thus left to see, after all, what {\it new} 
kind of `charges' may ever characterize a pair 
of mutually `charge conjugate' fermion fields 
each having only {\it two} (rather than 
four) freedom degrees, and what should be 
the {\it new} meaning to be assigned, accordingly,  
to each single relationship $\psi'(x) = 
U_C \psi'^{\dagger {\rm T}}(x)$ and $\psi'^c(x) 
= U_C \psi'^{c \dagger {\rm T}}(x)$.

\section{A spin-1/2 fermion with mass of the `Majorana' 
(rather than `Dirac') kind as a particle 
correspondingly endowed with pseudoscalar-type 
(rather than scalar-type) charges}
Let us look again at either (\ref{3.1}) or 
(\ref{3.2bis}). If we still assume the 
two fermion fields $\psi(x),\psi^c(x)$ $-$
characterized by Eq.~(\ref{2.9bis}) $-$ 
to be also mass eigenfields (as in the 
`Dirac mass' case), we are clearly left 
with the usual `charged' spin-$\frac{1}{2}$ 
particles. If we instead suppose the 
field pair $\psi'(x),\psi'^c(x)$ to be 
just an alternative pair of mass-eigenfields
(as in the `Majorana mass' case), we can no longer 
expect the associated fermions to be `really neutral' 
(as commonly believed), the reason being because 
$\psi'(x)$ and $\psi'^c(x)$ are themselves 
two `mutually charge-conjugate' (rather than 
individually self-conjugate) fields. In this 
case, as we have already pointed out, we 
should be actually faced with a {\it new} 
type of `charged' spin-$\frac{1}{2}$ particles.

	To see what basic peculiar features may 
truly distinguish the latter `charged' fermion type 
from the former one, it appears crucial to compare 
how, in the two cases, the single `particle' and 
`antiparticle' annihilation (creation) operators 
do transform under {\it space reflection}. 
In either case, each field that happens to be a mass 
eigenfield turns out to be also a solution of a 
Dirac-type equation. Thus, if Dirac invariance 
with respect to space reflection is always invoked, 
one gets e.g.
\begin{equation}
P\psi(x^R)P^{-1} = i \gamma^0 \psi(x),
\;\;
P\psi^c(x^R)P^{-1} = i \gamma^0 \psi^c(x)
\label{3.12ter}
\end{equation}

\noindent
for the `Dirac mass' case,
and 
\begin{equation}
P'\psi'(x^R)P'^{-1} \!=\! i \gamma^0 \psi'(x),
\;\;
P'\psi'^c(x^R)P'^{-1} \!=\! i \gamma^0 \psi'^c(x)
\label{3.12quater}
\end{equation}

\noindent
for the `Majorana mass' case, where 
$x^R \equiv(t,-\bf r)$ (and where the 
phase choice is such that it allows 
either $P$ or $P'$ to commute with $C$ 
and to imply, as required, a fermion--antifermion 
relative intrinsic parity $i^2=-1$). 
It can be argued that the two parity operators 
$P$ and $P'$ $-$ the former defined by 
(\ref{3.12ter}) and the latter by (\ref{3.12quater}) 
$-$ provide two {\it non}coinciding representations 
of $x \rightarrow x^R$ (just relevant to 
the two cases under consideration). That 
$P$ and $P'$ do {\it not} overlap can be 
easily shown by direct use of (\ref{3.1}) 
or (\ref{3.2bis}). Bearing in mind that 
either $P$ or $P'$ is (in itself) nothing 
but a linear operator acting on annihilation 
and creation operators, one should expect, 
for example, that $P'$ applied to (\ref{3.2bis}) 
still gives
\begin{equation}
\left\{
\begin{array}{lcl}
 P'\psi(x)P'^{-1}\!\!&=&\!\!X_{\rm L} P'\psi'(x)P'^{-1} + X_{\rm R} P'\psi'^c(x)P'^{-1} 
\\ [0.05in] 
P'\psi^c(x)P'^{-1}\!\!&=&\!\!X_{\rm R} P'\psi'(x)P'^{-1} + X_{\rm L} P'\psi'^c(x)P'^{-1}. 
\end{array}
\right. 
\label{3.1ter}
\end{equation}

\noindent
Hence, substituting (\ref{3.12quater}) (and recalling 
that $X_{\rm L,R}\gamma^0=\gamma^0X_{\rm R,L}$), 
one is actually led to a result which is 
{\it not} the same as (\ref{3.12ter}):
\begin{equation}
P'\psi(x^R)P'^{-1} = i \gamma^0 \psi^c(x),
\;\;
P'\psi^c(x^R)P'^{-1} = i \gamma^0 \psi(x).
\label{3.23bis}
\end{equation}

\noindent
If $\psi(x)$ and $\psi^c(x)$ are still mass eigenfields
(with usual normal mode expansions), then 
\begin{equation}
Pa^{(\mp)}({\bf p})P^{-1} = i a^{(\pm)}(-{\bf p})
\mbox{, and so on} \;\;\;\; \mbox{(`Dirac mass')};
\label{3.13bis}
\end{equation}
	
\noindent
which means, strictly speaking, that standard 
Dirac particles $-$ or `Dirac mass' fermions $-$ 
are bound to carry charges behaving just like 
ordinary {\it scalars}. This cannot be true for 
`Majorana mass' fermions, i.e. for the new kind 
of `charged' spin-$\frac{1}{2}$ particles associated 
with the alternative pair $\psi'(x),\psi'^c(x)$ 
of mass eigenfields. The fact is that, even though 
Eq.~(\ref{3.12quater}) is quite analogous to 
Eq.~(\ref{3.12ter}), the expansions strictly defining 
$\psi'(x)$ and $\psi'^c(x)$ as mass eigenfields 
are the {\it non}standard ones (\ref{2.35}) and 
(\ref{2.36}). So, space reflection now implies
\begin{equation}
P'a^{(\mp)}({\bf p})P'^{-1} = i b^{(\pm)}(-{\bf p})
\mbox{, and so on}\;\;\;\; \mbox{(`Majorana mass')},
\label{3.13ter}
\end{equation}
	
\noindent
thus acting on `Majorana mass' fermions {\it in the 
same way as the whole CP operation acts on `Dirac mass' 
fermions}. This makes sense, of course, if and only if 
`Majorana mass' fermions are assumed to carry nonzero 
charges behaving just like {\it pseudoscalars}. Due to 
such charges, one and the same particle of this kind 
is indeed predicted to behave  like {\it either} a 
`fermion' {\it or} an `antifermion' according 
to the given chirality involved, so that, in the 
ultrarelativistic limit, it would naturally approach 
an exact {\it two-component} particle model. Associated 
with it, there should also be a {\it non}vanishing 
(though not generally conserved) current, proportional 
e.g. to 
\begin{equation}
\bar{\psi'}\gamma^\mu(-\gamma^5)\psi' 
= \bar{\psi'}_{\rm L}\gamma^\mu\psi'_{\rm L} - 
\bar{\psi'}_{\rm R}\gamma^\mu\psi'_{\rm R}
\label{3.23}
\end{equation}

\noindent 
($\mu=0,1,2,3$) or to the current operator 
`charge conjugate' to (\ref{3.23}). For speed 
zero (or at least negligible compared with $c$) 
such a particle should be {\it equally} able 
to look like a `fermion' (with negative chirality) 
{\it as} like an `antifermion' (with positive chirality), 
while for ultrarelativistic speeds it should tend, 
depending on its helicity, to behave either 
in the former or in the latter manner {\it only}. 
Similarly, its individual helicity eigenstates 
could never turn out to be {\it sharp} 
`charge eigenstates' but could only tend to 
become so in the ultrarelativistic limit 
(or in the limit of zero mass $[{\ref{Barut1993}}]$).
A particle like this cannot be strictly said 
to be `chargeless' (as it would be for a 
{\it true} Majorana particle). The general fact 
is left, anyhow, that {\it whatever spin-$\frac{1}{2}$ 
particle endowed with `Majorana mass' would still 
be `neutral' with respect to scalar-type charges}.

	A comparison of Eqs.~(\ref{3.23bis}),(\ref{3.13ter}) 
with Eqs.~(\ref{3.12ter}),(\ref{3.13bis}) shows that 
actually,
\begin{equation}
P' = CP \;(= PC).
\label{3.25}
\end{equation}

\noindent
Such a relationship, just equating 
parity $P'$ for `Majorana mass' fermions 
with $CP$ for `Dirac mass' fermions, 
may in particular lead one to speculate 
that, if all particles of the latter type 
were replaced by particles of the former type, 
the usual $CP$ mirror symmetry of weak 
processes would then become nothing but 
a {\it genuine} (ordinary) mirror symmetry! 
In view of Eq.~(\ref{3.25}), one may also 
put, for convenience,
\begin{equation}
P = P'_{\rm ex} \Longrightarrow P' = CP'_{\rm ex}
\; (= P'_{\rm ex}C),
\label{3.26}
\end{equation}

\noindent
where $P'_{\rm ex}$ stands just for an `external' 
parity operator (identical with $P'$ except for 
not involving pseudoscalar-charge reversal). This 
shows that, on passing from the `Dirac mass' case 
(when $C$ may specifically be said to invert scalar-type 
charges) to the `Majorana mass' case (when $C$ may 
specifically be said to invert pseudoscalar-type charges), 
the standard effect of (maximum) $P$ violation 
would indeed be reduced to a mere effect of (maximum) 
$P'_{\rm ex}$ violation.

\section{`Scalar-charge conjugation' and 
`pseudoscalar-charge conjugation' operations}
	It follows from the foregoing that $C$ does in principle 
reverse {\it scalar-type } as well as {\it pseudoscalar-type} 
charges, giving rise always to a {\it full} particle 
$\rightleftharpoons$ antiparticle interchange. If so, 
how can one {\it separately} think of a `scalar-charge 
conjugation' and a `pseudoscalar-charge conjugation' 
operation? Let two such individual operations be 
denoted by $C_{\rm s}$ and $C_{\rm ps}$, respectively 
(with $C_{\rm s}^2=C_{\rm ps}^2=1$). Actually, 
it is only the {\it product} of them both, 
i.e. $C$ itself, that is strictly demanded to result 
in a {\it pure} operation acting on annihilation 
and creation operators as in (\ref{2.28}). It appears 
legitimate, therefore, to attempt to define these 
single charge-conjugation operations so that, 
regardless of the mass kind involved, they may 
simply fulfil the general requirements
\begin{equation}
C_{\rm s}\Psi(x)C_{\rm s}^{-1}
= C\Psi(x)C^{-1},
\;\;
C_{\rm s}\Psi'(x)C_{\rm s}^{-1}
= \Psi'(x)
\label{4.3}
\end{equation}

\noindent
and
\begin{equation}
C_{\rm ps}\Psi(x)C_{\rm ps}^{-1} = \Psi(x),
\;\;\;
C_{\rm ps}\Psi'(x)C_{\rm ps}^{-1}
= C\Psi'(x)C^{-1},
\label{4.3bis}
\end{equation}

\noindent
where $\Psi(x)$ and $\Psi'(x)$ stand for
the column matrices in Eqs.~(\ref{3.3ter}) 
and (\ref{3.3quater}), and where $C=
C_{\rm ps}C_{\rm s}$ ($=C_{\rm s}C_{\rm ps}$).
Definitions (\ref{4.3}) and (\ref{4.3bis}) 
naturally embody, in particular, 
the new achievement that `Majorana mass' 
eigenfields are themselves non-neutral 
fields associated with pseudoscalar-type 
(rather than scalar-type) charges. 
Making use of transformation (\ref{3.1}) 
or its inverse (\ref{3.2bis}), we can see 
that neither $C_{\rm s}$ nor $C_{\rm ps}$ 
as given above may be fundamentally 
represented as {\it purely} acting 
(like $C$) on annihilation and 
creation operators: for instance, 
it can be easily checked that, for 
the two requirements in (\ref{4.3}) 
to hold simultaneously, the $C_{\rm s}$ 
operation must be understood to be 
further {\it such that} 
\begin{equation}
\left\{ 
\begin{array}{lcr}
C_{\rm s}\left[
\frac{1}{2}(1 \mp \gamma^5)\psi\right]C_{\rm s}^{-1}
=\frac{1}{2}(1 \pm \gamma^5)C \psi C^{-1}
= U_C \left[ \frac{1}{2}(1 \mp \gamma^5)\psi \right]^{\dagger {\rm T}}
\\[.1in]  
C_{\rm s}\left[ \frac{1}{2}(1 \pm \gamma^5)\psi^c \right] C_{\rm s}^{-1}
=\frac{1}{2}(1 \mp \gamma^5)C \psi^c C^{-1}
= U_C \left[ \frac{1}{2}(1 \pm \gamma^5)\psi^c \right]^{\dagger {\rm T}}.
\label{3.22ter}
\end{array}
\right.
\end{equation} 

	Still focusing on (\ref{4.3}), 
let us compare $C_{\rm s}$ with $C$ 
as represented in Eqs.~(\ref{3.3}) and 
(\ref{3.10bis}). Due to (\ref{3.22ter}), 
we shall now have {\it not only}
\begin{equation}
C_{\rm s}\Psi(x)C_{\rm s}^{-1} 
= \pmatrix{U_C&0\cr 0&U_C\cr}
\pmatrix{\psi^{\dagger {\rm T}}(x)\cr \psi^{c \dagger {\rm T}}(x)\cr}
\label{4.4}
\end{equation}

\noindent
{\it but also}
\begin{equation}
C_{\rm s}\Psi'(x)C_{\rm s}^{-1} 
= \pmatrix{U_C&0\cr 0&U_C\cr}
\pmatrix{\psi'^{\dagger {\rm T}}(x)\cr \psi'^{c \dagger {\rm T}}(x)\cr}.
\label{4.5}
\end{equation}

\noindent
This leads us, in particular, to the conclusion 
that the conventional identity $C \psi'(x) C^{-1} = 
U_C \psi'^{\dagger {\rm T}}(x)$ is to be
properly recast {\it into} 
\begin{equation}
C_{\rm s} \psi'(x) C_{\rm s}^{-1}= U_C \psi'^{\dagger {\rm T}}(x).
\label{2.9'ter}
\end{equation}

	In the light of (\ref{3.22ter}) and (\ref{2.9'ter}),
it is immediate to see that $C_{\rm s}$ is exactly 
coinciding with the special `charge conjugation' 
operation $C'$ (distinct from $C$) mentioned in Sec.~3. 
As shown by (\ref{2.9'ter}), it is thus $C_{\rm s}$ 
that is normally {\it mistaken} for $C$ on dealing 
with `Majorana mass' fermions! This is a fundamental 
outcome which enables us to shed light, at last, 
on the {\it new} meaning to be assigned to the 
formal relationship $\psi' = U_C \psi'^{\dagger {\rm T}}$ 
(or its `charge conjugate' $\psi'^c = U_C \psi'^{c \dagger {\rm T}}$):
\bigskip 

{\it Strictly speaking, the constraint}  
$\psi' = U_C \psi'^{\dagger {\rm T}}$ ($\psi'^c 
= U_C \psi'^{c \dagger {\rm T}}$) 
{\it does only express neutrality of} 
$\psi'$ ($\psi'^c$) {\it with respect to 
scalar-type charges, and not real neutrality 
of} $\psi'$ ($\psi'^c$).
\bigskip
 	
	Bearing in mind transformation 
(\ref{3.1}), let us then consider an 
{\it active} `Majorana mass' neutrino, 
associated with a field of the $\psi'$-type, 
and a {\it sterile} one, associated 
with the partner field of the $\psi'^c$-type. 
Herein, two such `Majorana mass' 
neutrino versions are charge conjugate 
to each other (and no longer individually 
self-conjugate). This, however, does not 
exactly mean that they can now be said 
to represent {\it just} a `lepton' 
and {\it just} an `antilepton' (as it 
normally happens for a `Dirac mass' 
neutrino and the corresponding antineutrino). 
The point is that, unlike any standard 
neutrino--antineutrino pair, they 
would share a {\it pseudoscalar} 
(rather than scalar) `lepton number' 
variety (proportional to chirality). 
We thus have that one and the same 
{\it active} `Majorana mass' neutrino 
$-$ associated with a current of the 
type (\ref{3.23}) $-$ could in turn be 
{\it either} a left-handed `lepton' 
(with {\it positive} lepton number) 
{\it or} a right-handed `antilepton' 
(with {\it negative} lepton number), 
and the exact converse (with `lepton' 
and `antilepton' interchanged) would 
in principle apply to the (`charge 
conjugate') {\it sterile} version of it. 
Note, on the other hand, that due to the 
presence of mass (which breaks chirality 
conservation) the lepton number at issue 
may be conserved only in magnitude, 
and {\it not} in sign. Hence, with the 
help of (\ref{2.35}), it is in particular 
easy to realize that such a neutrino 
(taken in its active version) could 
invariably give rise to a net neutrinoless 
double $\beta$-decay {\it even without being 
a really neutral particle}. It is worth 
pointing out, moreover, that a given 
{\it left-handed} ({\it right-handed}) 
neutrino would remain coupled to a given 
{\it left-handed} `charged lepton' 
({\it right-handed} `charged antilepton'), 
so that we should always be able, after all, 
to recognize single lepton families marked 
by their own `lepton-number conserving' 
weak currents. This should be related to 
the general fact $-$ already emphasized 
in Sec.~4 $-$ that `Dirac mass' fermions 
themselves, when involved in weak processes, 
are apparently described by `dynamical 
eigenfields' structured just like
$\psi'(x)$ and $\psi'^c(x)$ (as if they 
were as well carrying a {\it pseudoscalar} 
charge variety which is normally kept 
`hidden' in their strict Dirac behaviors 
and may indeed be revealed once weak interaction 
is turned on $[\ref{Ziino2006}-\ref{Ziino2007}]$).

\section{Pseudoscalar-type charges and 
the $CPT$ theorem}
As is well-known, one familiar example 
of a pseudoscalar-type charge is just 
provided by {\it magnetic} charge 
$[{\ref{Barut1972}-\ref{Ziino2000}}]$. 
The fact that the field ${\bf H}$ generated 
by a magnetic dipole is an axial-vector 
(invariant under parity) clearly means that 
space reflection does also interchange 
the signs of the two poles (besides 
interchanging the spatial locations of 
them). The fact, on the other hand, that 
${\bf H}$ is inverted by time reversal 
(instead of being left unvaried 
like an electric field) shows that each 
pole does undergo once again a change 
in sign if space reflection is followed 
by time reversal. We therefore have that 
the overall effect of both space and time 
inversions on magnetic charge would be 
still the same as on electric charge. 
This holds as well for the whole $CPT$ 
operation, which would regularly turn 
a magnetic monopole with a given 
four-momentum into an {\it opposite} 
monopole with identical four-momentum. 
Such a conclusion (obviously extensible 
to whatever pseudoscalar-type charges) 
is also supported by the fact that 
the `proper' relativistic transformation 
of {\it strong reflection} $-$ essentially 
equivalent to $CPT$ [${\ref{Sakurai1964bis}}
-{\ref{Recami1976}}$] $-$ acts identically 
on {\it vector} as on {\it pseudovector} 
currents.

	That said, let us go back to 
the revised `Majorana mass' fermion model 
herein proposed. The basic peculiar 
feature of it is obviously given 
by (\ref{3.13ter}). In short, we have 
that the parity operator $P'$ (relevant 
to the `Majorana mass' case) does 
naturally exert a {\it `charge conjugating' 
internal extra action}, and this may 
clearly happen only for a particle just 
endowed with pseudoscalar-type charges. 
As already stressed in Sec.~4, 
such a result tells us that 
$P'$ is exactly the same as $CP$ 
for standard fermions. We also have, 
therefore, that the only symmetry 
under $CP$ ($=P'$) does not enable us 
to distinguish between scalar-type 
and pseudoscalar-type charges, 
and so it may {\it equally} allow, 
in principle, either the conjecture 
of a `Dirac mass' neutrino (endowed 
with a {\it scalar} lepton number) 
or the conjecture of a `Majorana mass' 
neutrino (endowed with a {\it pseudoscalar} 
lepton number). 
	
	To enlarge the discussion to 
$CPT$ symmetry, let us just denote by $T$ 
and $T'$ the two (antiunitary) operators 
representing time reversal in the pure 
`Dirac mass' and `Majorana mass' 
cases, respectively. In the light of 
the above remarks on the behaviors 
of pseudoscalar-type charges, such 
operators cannot be expected to coincide, 
the reason being because $T'$ is also 
demanded (just like $P'$) to include 
 a {\it `charge conjugating' internal 
effect}. This can be properly obtained 
in terms of `particle' and `antiparticle' 
annihilation (creation) operators such as 
$a^{(h)}({\bf p})$ ($a^{(h)\dagger}({\bf p})$) 
and $b^{(h)}({\bf p})$ ($b^{(h)\dagger}({\bf p})$). 
In other words, considering that (except 
for phase factors)
\begin{equation}
Ta^{(h)}({\bf p})T^{-1} = a^{(h)}(-{\bf p})
\mbox{, and so on} \;\;\;\;\;\; \mbox{(`Dirac mass')},
\label{4.9bis}
\end{equation}
	
\noindent
it should correspondingly result
\begin{equation}
T'a^{(h)}({\bf p})T'^{-1} = b^{(h)}(-{\bf p})
\mbox{, and so on} \;\;\;\;\;\; \mbox{(`Majorana mass')}.
\label{4.9ter}
\end{equation}
	
\noindent
We thus have that $T'$ can be formally related 
to $T$ as follows:
\begin{equation}
T' = CT,
\label{4.9quater}
\end{equation}

\noindent
where, in analogy with (\ref{3.26}), it may be set
\begin{equation}
T = T'_{\rm ex} \Longrightarrow T' = CT'_{\rm ex},
\label{4.14}
\end{equation}

\noindent
with $T'_{\rm ex}$ standing for a mere `external' 
time-reversal operator (identical with $T'$ except 
for not involving pseudoscalar-charge reversal).
From (\ref{3.25}) and (\ref{4.9quater}), on the 
other hand, it can be seen that the overall $PT$ 
and $P'T'$ operators do indeed {\it coincide}. 
Hence, 
\begin{equation}
CPT  = CP'T' = CP'_{\rm ex}T'_{\rm ex},
\label{4.2quater}
\end{equation}

\noindent
and this is in full accordance with the 
above-mentioned fact that the whole 
(equivalent) symmetry operation of 
{\it strong reflection} does not make
any distinction between scalar-type 
and pseudoscalar-type charges.

	It may therefore be concluded that 
the $CPT$ theorem is just as well available 
for the special new kind of {\it charged} 
particles to which `Majorana mass' fermions 
should strictly correspond. This, however, 
does not really mean that an active field 
$\psi'$ ($=X_{\rm L}\psi + X_{\rm R}\psi^c$) 
and its sterile counterpart $\psi'^c$ 
($=X_{\rm L}\psi^c + X_{\rm R}\psi$) 
are bound to have {\it identical} masses 
as a result of their being also a pair of 
`mutually charge-conjugate' fields. If $\psi'$ 
is coupled, say, to a mass $m_{\rm L}$, and 
$\psi'^c$, say, to a mass $m_{\rm R}$, we may,
in other words, generally assume $m_{\rm L}\not=m_{\rm R}$ 
as in the conventional approach. The reason 
is just because, for fields having such 
expansions as (\ref{2.35}) and (\ref{2.36}), 
the whole symmetry operation (\ref{4.2quater}) 
is only able (like $P'$ and unlike $C$) 
to connect annihilation or creation operators 
always included in {\it one and the same} 
expansion. As an immediate consequence, 
one has e.g. that the See--Saw Mechanism 
may still apply to neutrino masses without 
spoiling $CPT$ symmetry.

	Yet, there seems to be another 
intriguing feature to be pointed out. 
According to the new approach, fermions 
with `Majorana masses' would indeed obey 
{\it ordinary} mirror symmetry (as just 
the {\it analogue} of $CP$ symmetry 
for standard fermions) and they would also 
experience a {\it manifest} (maximum) 
$C$ violation (due to the fact that 
an {\it active} `Majorana mass' fermion 
and its {\it sterile} counterpart are now 
`charge conjugate to each other'). 
Hence, neglecting the extreme conjecture 
of $CPT$ breakdown $[{\ref{Colladay1998}-\ref{Esposito2010}}]$, 
we see that such fermions would as well 
obey symmetry under $CT'$ ($=T'_{\rm ex}$), 
even though at the price of also 
experiencing a (maximum) {\it time reversal} 
(i.e. $T'$) violation, which would just 
counterbalance the `recovered' ordinary 
mirror symmetry. In close connection 
with this, it is worth noting that 
merely releasing the constraint 
$m_{\rm L}=m_{\rm R}$ would already
break $C$ and $T'$ individual symmetries, 
with no need to consider weak dynamics. 
We thus have, for example, that active 
and sterile `Majorana mass' neutrino 
versions which are supposed to have masses 
$m_{\rm L}\not=m_{\rm R}$ should anynow 
be taken as particles {\it intrinsically 
violating} either $C$ or time reversal.

\section{Single `Dirac mass' or `Majorana mass' 
charged fermion fields as superpositions of 
pure Majorana neutral fields}
	We know from Ref.~$[{\ref{Majorana1937}}]$
$-$ see also Refs.~$[{\ref{Bilenky1987}},{\ref{Giunti2007}}]$ $-$ 
not only that a single Majorana neutral 
field $\psi_{\rm M}(x)$ can be obtained, 
via Eq.~(\ref{2.1}), as a superposition of 
two distinct (and mutually charge-conjugate)
`Dirac mass' charged fermion fields, but even 
that a single `Dirac mass' charged fermion field 
$\psi(x)$ may be seen, conversely, as a 
superposition of two distinct Majorana
neutral fields (with opposite $CP$ intrinsic 
parities). One has, more precisely,
\begin{equation}
\psi(x) = \frac{1}{\sqrt{2}}\left[\psi^{(+)}_{\rm M}(x) 
+ i\psi^{(-)}_{\rm M}(x)\right] 
\;\;\;\;\;\; \mbox{(`Dirac mass')},
\label{8.1}
\end{equation}

\noindent
where $\psi^{(+)}_{\rm M}(x)=\psi_{\rm M}(x)$,
and where $\psi^{(-)}_{\rm M}(x)$ is a new 
(still manifestly self-conjugate) Majorana field 
defined as
\begin{equation}
\psi^{(-)}_{\rm M}(x) = \frac{-i}{\sqrt{2}}\left[\psi(x) 
- \psi^c(x)\right].
\label{8.2}
\end{equation}

\noindent
The superscripts $(\pm)$ have here been used 
to denote the $CP$ intrinsic parities 
distinguishing two such neutral fields. 
A similar result holds for the `Dirac mass' 
field being the charge conjugate of $\psi(x)$, 
which may likewise be expressed in the form
\begin{equation}
\psi^c(x) = \frac{1}{\sqrt{2}}\left[\psi^{(+)}_{\rm M}(x) 
- i\psi^{(-)}_{\rm M}(x)\right] 
\;\;\;\;\;\; \mbox{(`Dirac mass')}.
\label{8.3}
\end{equation}

\noindent
Whether we are considering (\ref{8.1})
or (\ref{8.3}), fields $\psi_{\rm M}^{(\pm)}(x)$
are understood to be mass eigenfields with
identical eigenvalues, and their masses 
(the same as those carried by $\psi$ and 
$\psi^c$) may be said to display `Dirac-like' 
characters.

	In the light of transformation 
(\ref{3.2bis}) $-$ which turns Eq.~(\ref{2.1})
into Eq.~(\ref{2.34}) $-$ we may now, 
on the other hand, also think 
of a Majorana neutral field being 
a superposition of two mutually 
charge-conjugate `Majorana mass' 
charged fermion fields, $\psi'(x)$ and 
$\psi'^c(x)$, with identical masses. 
This, indeed, is in line with the general 
fact that, due to condition (\ref{2.2bis}), 
the mass term relevant to a true Majorana 
field may be claimed to be {\it equally 
reminiscent} of a `Dirac' as of a 
`Majorana' mass term. Starting from 
$\psi'(x)$ and $\psi'^c(x)$ (with masses 
$m_{\rm L}$ and $m_{\rm R}$ being 
in particular such that $m_{\rm L}
=m_{\rm R}$), we can, thus, again 
construct two independent (manifestly 
self-conjugate) Majorana neutral fields 
as above. They read
\begin{equation}
\psi'^{(+)}_{\rm M}(x) = \frac{1}{\sqrt{2}}\left[\psi'(x) + \psi'^c(x)\right],
\;\;\;
\psi'^{(-)}_{\rm M}(x) = \frac{-i}{\sqrt{2}}\left[\psi'(x) - \psi'^c(x)\right]
\label{8.4}
\end{equation}

\noindent
and still have opposite $CP$ intrinsic 
parities. Hence we can see that, in 
full analogy with the `Dirac mass' case, 
fields $\psi'(x)$ and $\psi'^c(x)$ 
themselves (taken with identical masses) 
may be split, conversely, as follows:
\begin{equation}
\left\{
\begin{array}{lcl}
\!\psi'(x) \!\!\!&=&\!\!\! \frac{1}{\sqrt{2}}\left[\psi'^{(+)}_{\rm M}(x) 
+ i\psi'^{(-)}_{\rm M}(x)\right] 
\\ [0.09in] 
\!\psi'^c(x) \!\!\!&=&\!\!\! \frac{1}{\sqrt{2}}\left[\psi'^{(+)}_{\rm M}(x) 
- i\psi'^{(-)}_{\rm M}(x)\right] 
\end{array}
\;
\mbox{(`Majorana mass'\,; $m_{\rm L}=m_{\rm R}$)}.
\right.
\label{8.5}
\end{equation}

\noindent
Of course, the two neutral fields $\psi'^{(\pm)}_{\rm M}(x)$ 
in (\ref{8.5}) may correspondingly be said to have masses 
displaying `Majorana-like' characters.

\section{Concluding remarks}
	There are mainly two motivations
underlying this paper. The former one
is the purpose of throwing light upon 
some basic theoretical inconsistencies
which appear to be present in the usual 
approach to Majorana fermions and 
`Majorana mass' fermions. The latter one 
is the need of working out accordingly 
$-$ with no departure from standard QFT 
$-$ a formalism being really free of 
such inconsistencies and being further 
able to lead, after all, to a new insight 
into the whole subject.

	As a starting point, a brief 
discussion has been made on how 
to interpret the two couples of massless 
Dirac field solutions (\ref{2.10}) and (\ref{2.11}) 
in order to avoid that charge conjugation 
$C$ may happen to invert helicities. 
It has been argued that the appropriate 
reading is provided by (\ref{2.10ter}) 
and (\ref{2.11ter}), and {\it not} by 
(\ref{2.10bis}) and (\ref{2.11bis}), 
even though the latter choice seems just 
to come from a natural extension of the 
well-known Dirac `prescription' (\ref{2.9bis}). 
Opting for (\ref{2.10ter}) and (\ref{2.11ter}) 
has also been shown to be the only choice 
that correctly implies $C$ {\it violation} as a 
result of any {\it asymmetry} occurring 
between (\ref{2.10}) and (\ref{2.11}). 
The fact is left, however, that the 
alternative reading is also the one which 
normally allows a `Majorana mass' neutrino 
to be recognized as a {\it self-conjugate} 
particle! What, then, about the real nature 
of such a neutrino? To shed full light 
on the matter, a truly {\it direct} (and 
thus unambiguous) check has been tried, 
based on the {\it primary} QFT definition 
itself of charge conjugation, i.e. its 
{\it fundamental representation} (\ref{2.28}) 
(in the Fock space). This procedure has led 
us to conclude that {\it a `Majorana mass' 
neutrino, unlike the original neutrino by 
Majorana himself, cannot be really claimed 
to be genuinely self-conjugate}. The point 
may be generally set as follows. Take the 
wholly {\it active} ({\it sterile}) fermion 
field which can be obtained from suitably 
mixing the chiral components of two mutually 
charge-conjugate Dirac fields. If $C$ is 
{\it exactly} applied as in (\ref{2.28}), 
the net outcome is that such a field is 
correspondingly turned into its {\it sterile} 
({\it active}) counterpart, and {\it not} 
into itself! This also shows that 
the standard formula (\ref{2.9bis}) $-$ 
just suitable for defining the charge conjugate 
of a `Dirac mass' fermion field $-$ cannot 
be extended to `Majorana mass' fermion fields 
(and thus be used as `proof' of their real 
neutrality) without coming into conflict 
with the true $C$ definition (\ref{2.28}). 

	If so, how can we explain the 
individual constraints, $\psi'(x) 
= U_C \psi'^{\dagger {\rm T}}(x)$ and
$\psi'^c(x) = U_C \psi'^{c\dagger {\rm T}}(x)$, 
naturally applying to an active 
`Majorana mass' fermion field $\psi'(x)$ 
and its (`charge conjugate') sterile version 
$\psi'^c(x)$? The answer to this crucial 
question can actually be found in the 
{\it new} formalism which has been herein 
developed to remodel a fermion with mass 
of the `Majorana' (rather than `Dirac') type. 
The basic novelty is that replacing 
a `Dirac mass' with a `Majorana mass' 
does now mean passing from a standard fermion, 
endowed with {\it scalar-type} charges, to a 
fermion which is {\it not} really neutral 
but is endowed with {\it pseudoscalar-type} 
charges. A `charged' spin-$\frac{1}{2}$ 
particle like this should essentially be 
thought of as a {\it chiral} object 
in turn behaving like a `fermion' or 
an 'antifermion' according to {\it either} 
sign of the associated chirality (where 
`fermion' and `antifermion' should clearly 
appear {\it interchanged} for the particle 
`charge conjugate' to it). In such a framework, 
charge conjugation $C$ proves to act as a 
{\it true} `particle--antiparticle conjugation' 
operation which may generally be split 
into the product of a mere `scalar-charge 
conjugation' and a mere `pseudoscalar-charge 
conjugation' operation. Hence it can indeed 
be seen that the above constraints peculiar 
to $\psi'(x)$ and $\psi'^c(x)$ are just  
expressing {\it neutrality of `Majorana mass' 
fermions with respect to scalar-type charges}. 

	This, however, does {\it not} mean 
that a genuinely neutral Majorana fermion 
cannot exist in nature. Such a particle 
$-$ strictly described by a {\it manifestly 
self-conjugate} field like the one given 
in Eq.~(\ref{2.1}) or (\ref{2.34}) $-$
would regularly possess {\it no charges 
at all} (i.e. neither {\it scalar-type} 
nor {\it pseudoscalar-type} charges). 
Herein a `new' model of it has been 
obtained, where some inconsistent features 
unavoidably affecting the conventional 
theory appear to be automatically removed. 
Firstly, in full compliance with Eq.~(\ref{2.1ter}), 
one now has that a genuine Majorana fermion 
can no longer be assigned any special sort 
of `handedness' marking it as {\it just} 
an `active' or `sterile' fermion:  there may 
always be only a fifty-fifty probability for it 
to look like the former or the latter particle. 
This implies, for example, that a true 
Majorana neutrino cannot really be claimed 
to be quite compatible with the SM (contrary 
to what may clearly be said for a {\it pure} 
active `Majorana mass' neutrino). Secondly, 
as {\it rigorously} demanded by the natural 
constraint (\ref{2.2bis}), the mass of a 
genuine Majorana fermion is now bound to be 
of a {\it single} kind, equally reminiscent 
of a `Majorana' as of a  `Dirac' mass kind. 

	The (no longer chargeless) `Majorana mass' 
fermion model here proposed does actually introduce 
a {\it new} sort of spin-$\frac{1}{2}$ particle 
which is somehow {\it half way} between the 
Dirac one and the genuine Majorana one. The point 
is that a `Majorana mass' fermion is now a particle 
endowed with {\it pseudoscalar-type} charges 
and still devoid of {\it scalar-type} charges: 
thus, it is `charged' (like a Dirac particle) 
but it also retains only {\it two} freedom degrees 
(like a true Majorana particle). This model 
applies, in principle, to any {\it wholly} 
active or {\it wholly} sterile massive fermions 
(including SUSY ones) which are usually (and 
improperly) referred to as `Majorana fermions' 
{\it tout court}. It in particular deals with 
the active and sterile versions of a `Majorana mass' 
neutrino as a pair of `mutually charge-conjugate' 
(rather than individually self-conjugate) 
particles which may always have two distinct 
mass values {\it all the same} (though now 
at the price of {\it intrinsically violating} 
$C$). The latter feature (still permitting a 
mass-generating mechanism like the See--Saw one) 
is just due to the following reason: unlike what 
happens for standard fermions (endowed with 
scalar-type charges), such (active and sterile) 
charge-conjugate neutrinos would {\it not} 
be really interchanged under $CPT$ (were it 
not so, on the other hand, $CPT$ itself would 
then be maximally violated!). Similarly, 
although a `Majorana mass' neutrino is now 
predicted to bear a {\it non}zero lepton number, 
we have that the conventional expectation 
for a neutrinoless double $\beta$-decay 
is left unaffected. This is simply because 
a {\it pseudoscalar} lepton number is as well
a quantity that changes sign {\it along with 
chirality}. Yet, supposing that real neutrinos 
should truly turn out to be `Majorana mass' 
(and not `Dirac mass') fermions, we have also 
that their well-known  phenomenology should 
then be reread in a way {\it opposite} to 
the usual one: the actual behaviors of them 
under space reflection and time inversion 
would indeed appear to have {\it reversed} 
meanings! The fact is that real neutrinos 
themselves should be admitted, accordingly, 
to be particles carrying {\it pseudoscalar-type} 
(rather than scalar-type) charges; so they 
would paradoxically seem to obey {\it pure} 
mirror symmetry (as just the {\it analogue} 
of `$CP$ symmetry' for standard fermions) 
and to violate instead (to a maximal degree) 
either {\it time reversal} or particle--antiparticle 
conjugation $C$ (with possible far-reaching 
effects on the yet unsolved `time arrow' 
fundamental question). 

	The last comment to be made 
goes beyond neutrino physics and is 
more generally addressed to the 
correspondence that has been herein 
set up between `Majorana masses' and 
pseudoscalar-type charges. Since one 
has that an active `Majorana mass' 
neutrino, whether viewed with {\it no} 
lepton number or with a {\it pseudoscalar} 
nonzero lepton number, is identically 
able to induce a net neutrinoless double 
$\beta$-decay, one could get the general 
idea that remodelling (active and sterile) 
`Majorana mass' fermions as particles 
endowed with pseudoscalar-type charges 
(rather than genuinely neutral) should 
truly have no direct repercussions in 
experimental reality. Such an idea, 
however, would leave out of account 
the fact that pseudoscalar-type charges 
could themselves be at the origin of 
{\it new} interactions. A particularly 
significant example may come from 
considering magnetic charge, whose 
pseudoscalar nature (opposed to the 
scalar nature of electric charge) 
is already well-known. This indeed 
suggests that `Majorana mass' fermions 
(just opposed to `Dirac mass' ones) 
might now be seriously expected to be 
even the natural candidates for magnetic 
{\it monopoles}.

\section*{Acknowledgments}
The author is particularly grateful to Dr. Salvatore 
Esposito for his incisive and detailed comments 
and his precious advice. Thanks are also due 
to Prof. V. V. Dvoeglazov, Dr. M. Dvornikov, 
Prof. E. Fiordilino and Dr. R. Plaga for useful 
suggestions.

\section*{References}
\begin{enumerate}
\item \label{Majorana1937} E. Majorana, {\it Nuovo Cimento} {\bf 14}, 171 (1937).
\item \label{Jehle1949} H. Jehle, {\it Phys. Rev.} {\bf 75}, 1609 (1949).
\item \label{Serpe1949} J. Serpe, {\it Phys. Rev.} {\bf 76}, 1538 (1949).
\item \label{Esposito1998} See e.g. S. Esposito, {\it Int. J. Mod. Phys. A} 
{\bf 13}, 5023 (1998); S. Esposito and N. Tancredi, {\it Eur. Phys. J. C} 
{\bf 4}, 221 (1998).
\item \label{Gell-Mann1979} M. Gell-Mann, P. Ramond, and R. Slansky, in
{\it Supergravity}, eds. D. Z. Freeman and P. van Nieuwenhuizen (North-Holland, Amsterdam, 1979).
\item \label{Mourik2012} V. Mourik, K. Zuo, S. M. Frolov, S. R. Brissard,
E. P. A. M. Bakkers, L. P. Kouwenhoven {\it Science} {\bf 336}, 1003 (2012).
\item \label{Brouwer2012} P. W. Brouwer, {\it Science} {\bf 336}, 989 (2012).
\item \label{Williams2012} J. R. Williams, A. J. Bestwick, P. Gallagher, Seung Sae Hong, Y. Cui, A. S. Bleich, J. G. Analytis, I. R. Fisher, and D. Goldhaber-Gordon, {\it Phys. Rev. Lett.} {\bf 109}, 056803 (2012).
\item \label{Rokhinson2012} L. P. Rokhinson, X. Liu and J. K. Furdyna, {\it Nature Physics} {\bf 8}, 795 (2012).
\item \label{Das2012} A. Das, Y. Ronen,	Y. Most, Y. Oreg, M. Heiblum and H. Shtrikman, {\it Nature Physics} {\bf 8}, 887 (2012).
\item \label{Deng2012} M. T. Deng, C. L. Yu, G. Y. Huang, M. Larsson, P. Caroff and H. Q. Xu, {\it Nano Lett.} {\bf 12}, 6414 (2012).
\item \label{Esposito2013} S. Esposito, {\it Europhys. Lett.} {\bf 102}, 17006 (2013).
\item \label{Merzbacher1970} See e.g. (no matter for some different notations therein used): E. Merzbacher, {\it Quantum Mechanics} (second edition) (John Wiley and Sons, New York, 1970) pp.~584,585.
\item \label{Dvornikov2012} M. Dvornikov, {\it Found. Phys.} {\bf 42}, 1469 (2012) (arXiv:1106.3303 {\bf [hep-th]}).
\item \label{Dvoeglazov2012} V. V. Dvoeglazov, {\it J. Phys: Conf. Series}
{\bf 343}, 012033 (2012).
\item \label{Glashow1961} S. L. Glashow, {\it Nucl. Phys.} {\bf 22}, 579 (1961).
\item \label{Weinberg1967} S. Weinberg, {\it Phys. Rev. Lett.} {\bf 19}, 1264 (1967).
\item \label{Salam1968} A. Salam, in {\it Proceedings of the Eighth Nobel Symposium on Elementary Particle Theory}, ed. N. Svartholm (Almquist and Wiksell, Stockholm, 1968) p.~367.
\item \label{Bilenky1987} S. M. Bilenky and S. T. Petcov, {\it Rev. Mod. Phys.} {\bf 59} (1987) 671.
\item \label{Giunti2007} C. Giunti and C. W. Kim, {\it Fundamentals of Neutrino Physics and Astrophysics} (Oxford University Press, 2007).
\item \label{McLennan1957} J. A. McLennan, Jr., {\it Phys. Rev.} {\bf 106}, 821 (1957).
\item \label{Case1957} K. M. Case, {\it Phys. Rev.}{\bf 107}, 307 (1957).
\item \label{Feynman1958} R. P. Feynman and M. Gell-Mann, {\it Phys. Rev.}
{\bf 109}, 193 (1958).
\item \label{Sudarshan1958} R. E. Marshak and E. C. G. Sudarshan, {\it Phys.
Rev.} {\bf 109}, 1860 (1958).
\item \label{Sakurai1958} J. J. Sakurai, {\it Nuovo Cimento} {\bf 7}, 649
(1958). 
\item \label{Sakurai1964} See e.g. J. J. Sakurai, {\it Invariance Principles and Elementary Particles} (Princeton University Press, Princeton, 1964) pp.~122,129--132.
\item \label{Dvoeglazov1997} V. V. Dvoeglazov, {\it Mod. Phys. Lett. A} {\bf 12}, 2741 (1997).
\item \label{Itzykson1985} C. Itzykson and J. Zuber, {\it Quantum Field Theor} 
(McGraw Hill, New York, 1985) pp.~87--89.
\item \label{Barut1993} A. O. Barut and G. Ziino, {\it Mod. Phys. Lett. A} 
{\bf 8}, 1011 (1993).
\item \label{Ziino2006} G. Ziino, {\it Int. J. Theor. Phys.} {\bf 45}, 1993 (2006). 
\item \label{Ziino2006bis} G. Ziino, {\it Ann. Fond. Louis de Broglie} {\bf 31}, 169 (2006).
\item \label{Ziino2007} G. Ziino, {\it Mod. Phys. Lett. A} {\bf 22}, 853 (2007).
\item \label{Barut1972} A. O. Barut, {\it Phys. Lett. B} {\bf 38}, 97 (1972);
{\bf 46}, 81 (1973).
\item \label{Defaria-Rosa1986} M. A. Defaria-Rosa, E. Recami, and W. A. 
Rodriguez Jr., {\it Phys. Lett. B} {\bf 173}, 233 (1986).
\item \label{Ziino1996} G. Ziino, {\it Int. J. Mod. Phys. A} {\bf 11}, 2081 (1996).	
\item \label{Ziino2000} G. Ziino, {\it Int. J. Theor. Phys.} {\bf 39}, 2605 (2000). 
\item \label{Sakurai1964bis} J. J. Sakurai, {\it Invariance Principles and 
Elementary Particles} (Princeton University Press, Princeton, 1964) pp.~136--143.
\item \label{Mignani1974} R. Mignani and E. Recami, {\it Nuovo Cimento A} {\bf 24},
438 (1974). 
\item \label{Mignani1975} R. Mignani and E. Recami, {\it Int. J. Theor. Phys.} 
{\bf 12}, 299 (1975).
\item \label{Recami1976} E. Recami and G. Ziino, {\it Nuovo Cimento A} 
{\bf 33}, 205 (1976).
\item \label{Colladay1998} D. Colladay and V. A. Kostelecky, {\it Phy. Rev. D} 
{\bf 55}, 6760 (1997); ibid. {\it D} {\bf 58}, 116002 (1998).
\item \label{Kostelecky2004} V. A. Kostelecky and M: Mewes, {\it Phy. Rev. D} 
{\bf 69}, 016005 (2004).
\item \label{Esposito2010} S. Esposito and G. Salesi, {\it Mod. Phys. Lett. A} 
{\bf 25}, 597 (2010).	

\end{enumerate}

\end{document}